\documentclass[reprint,amsmath,amssymb,aps,prb,groupedaddress,nofootinbib,twocolumn,superscriptaddress]{revtex4-2}
\usepackage{graphicx}
\usepackage{amsthm,amssymb,amsmath,braket,mathdots}
\usepackage{bm}
\usepackage[pagebackref=false,pdfnewwindow=true]{hyperref} %\hypersetup{draft}
\usepackage{epstopdf,psfrag}
\usepackage{relsize,amsbsy}
\usepackage[export]{adjustbox}

\usepackage{graphicx,xcolor} %tikz
\usepackage{tabularx, booktabs} %for tables with width

\usepackage{units} %\nicefrac,units
\usepackage{dsfont} %write unity 1

\DeclareMathOperator{\tr}{Tr}
\newcommand{\ii}{{\mathrm{i}}} %imaginary unit is straight
\newcommand{\e}{{\mathrm{e}}} %Euler's number straight
\renewcommand{\v}[1]{\bm{#1}} %write vectors fat

\renewcommand{\d}{\mathrm{d}}
\newcommand{\trans}{\mathrm{T}}
\newcommand{\ua}{{\uparrow}} %spin up
\newcommand{\da}{{\downarrow}} %spin down

\newcommand{\mac}{\mathcal}

\renewcommand{\eqref}[1]{Eq.~(\ref{#1})}

\definecolor{bananayellow}{rgb}{1.0, 0.88, 0.21}
\definecolor{straw}{rgb}{0.32, 0.28, 0.1}

\usepackage[caption=false]{subfig}
\usepackage{changes}
\definechangesauthor[name=Valentin: System size corrections, color=blue]{Vss} %for corrections refering to the fact that system size is the problem not the infite interaction
\definechangesauthor[name=Valentin: Corrections for right delta function, color=red]{Vdel} %for corrections refering to the new delta function
\definechangesauthor[name=Valentin: ToFinish, color=orange]{VL} %for corrections that need to be finished

\begin{document}
%\title{Landau level repulsion in a doped Mott insulator:\\ Quantum oscillations beyond the Onsager relation}
\title{Quantum oscillations in a doped Mott insulator beyond Onsager's relation}
\author{Valentin Leeb}
%\affiliation{Department of Physics TQM, Technische Universit{\"a}t M{\"u}nchen, James-Franck-Stra{\ss}e 1, D-85748 Garching, Germany}
\affiliation{Technical University of Munich, Germany; TUM School of Natural Sciences, Department of Physics, TQM}
\affiliation{Munich Center for Quantum Science and Technology (MCQST), 80799 Munich, Germany}
\author{Johannes Knolle}
 	%\affiliation{Department of Physics TQM, Technische Universit{\"a}t M{\"u}nchen, James-Franck-Stra{\ss}e 1, D-85748 Garching, Germany}
 	\affiliation{Technical University of Munich, Germany; TUM School of Natural Sciences, Department of Physics, TQM}
 	\affiliation{Munich Center for Quantum Science and Technology (MCQST), 80799 Munich, Germany}
 	\affiliation{\small Blackett Laboratory, Imperial College London, London SW7 2AZ, United Kingdom}
\date{\today}

\begin{abstract}
The kinetic energy of electrons in a magnetic field is quenched resulting in a discrete set of highly degenerate Landau levels (LL) which gives rise to fascinating phenomena like the de Haas–van Alphen effect (dHvAe) or the integer and fractional quantum Hall effects. The latter is a result of interactions partially lifting the degeneracy within a given LL while inter-LL interactions are usually assumed to be unimportant.  Here, we study the LL spectrum of the Hatsugai--Kohmoto model, a Hubbard-like model which is exactly soluble on account of infinite range interactions. For the doped Mott insulator phase in a magnetic field we find that the degeneracy of LLs is preserved but inter-LL interactions are important leading to a non-monotonous reconstruction of the spectrum. As a result, strong LL repulsion leads to aperiodic quantum oscillations of the dHvAe in contrast to Onsager's famous relation connecting oscillation frequencies with the Fermi surface areas at zero field. In addition, we find unconventional temperature dependencies of quantum oscillations and interaction-induced effective mass renormalizations. We discuss the general importance of inter-LL interactions for understanding doped Mott insulators in magnetic fields.
\end{abstract}
	
\maketitle

\section{Introduction}
The most remarkable aspect of Landau level (LL) formation of electrons in a magnetic field is the quenching of kinetic energy from a continuous spectrum to a set of discrete values. The resulting macroscopically large degeneracy lies at the heart of prominent effects like the integer quantum Hall effect (IQHE) \cite{Klitzing1980}, discovered 1980, as well as the dHvAe  already measured 50 years earlier \cite{deHaas1930}. There, the discreteness of the LL spectrum leads to quantum oscillations (QO) of thermodynamic and transport properties periodic in the inverse of the applied field~\cite{Shoenberg}. A natural and persistent research question then addresses  the role of electron interactions on the stability of the LL degeneracies and on physical observables.

The study of strong electron-electron correlations in orbital magnetic fields typically focuses on the single LL limit~\cite{fukuyama1979two,jain2007composite} because in the high magnetic field regime the spacing between LLs, i.e. the cyclotron frequency $\omega_c$, is large compared to the energy scale of the interactions. Prominently, it is well known that interactions in low LLs lead to a partial lifting of the LL degeneracy giving rise to the fractional quantum Hall effect (FQHE) \cite{Tsui1982,laughlin1983anomalous} in two-dimensions. However, the effect of LL formation is not constrained to two-dimensional systems nor to high magnetic fields where only very few LLs are occupied. For instance, QOs are routinely observed at much smaller fields in a huge variety of two- and three-dimensional materials, from weakly~\cite{Shoenberg} to strongly interacting ones~\cite{sebastian2012towards}, which calls for an investigation of strong correlation effects beyond the few LL limit. 

The effect of weak interactions on LLs is well understood within Fermi liquid theory and the semiclassical description of electron motion. At zero magnetic field effective single particle theories emerge as low-energy descriptions with renormalized parameters. In 1952 Onsager shaped our understanding of Fermi liquids in magnetic fields by a semiclassical picture \cite{Onsager1952}: The electrons perform  quantized orbital motion with the cyclotron frequency $\omega_c$, constrained by their energy-momentum dispersion~$\epsilon_{\v{k}}$ perpendicular to the magnetic field. This leads to Onsager's famous relation: The area of the extremal orbits around the Fermi surface equal the QO frequency. Note that these also determine the critical fields of the IQHE transitions in two dimension. The standard theory of QO was then completed by Lifshitz and Kosevich who connected the cyclotron frequency, which is determined by the effective mass $\omega_c = \nicefrac{e B}{m}$, to the universal temperature decay of the QO amplitude~\cite{Lifshitz1956}.

It is surprising that the canonical Onsager and Lifshitz--Kosevich (LK) theory, which is essentially a single particle theory, can be applied routinely even to strongly correlated systems like heavy fermion systems \cite{Taillefer1987} or cuprate high temperature superconductors \cite{doiron2007quantum,sebastian2012towards}. Nevertheless, in recent years numerous experimental findings \cite{tan2015unconventional,hartstein2018fermi,liu2018fermi,xiang2018quantum,pezzini2018unconventional,li2020emergent,leeb2021anomalous,czajka2021oscillations} have shown deviations to the standard theory of QOs. However, despite a number of effective theories available \cite{knolle2015quantum,zhang2016quantum,knolle2017excitons,sodemann2018quantum,erten2016kondo,Chowdhury2018,shen2018quantum,lee2021quantum} a controlled calculation including strong correlations is missing. 

Exactly soluble models have played an important role for understanding the physics of strongly correlated systems. Many phenomena, for example LL physics or gapless quantum spin liquid phases only emerge for large system sizes, which are challenging for numerical methods. However, in certain soluble limits rigorous progress can be made albeit with the trade-off of a fine-tuned set of parameters~\cite{moessner1996exact,kitaev2006anyons} or unphysical interactions~\cite{trugman1985exact,rokhsar1988superconductivity}. Important developments for understanding correlated electrons have been the dynamical mean field theory (DMFT), which is exact in infinite dimension, or the strongly coupled Sachdev--Ye--Kitaev models, which achieve exact solubility by random all-to-all couplings \cite{Sachdev1993,Kitaev2015,chowdhury2022sachdev}. Both limits have recently been extended to orbital magnetic field regimes and feature anomalous QOs~\cite{Chowdhury2018,acheche2017orbital,vuvcivcevic2021universal,vuvcivcevic2021electrical}.

Here, we concentrate on the Hatsugai--Kohmoto (HK) model, which is exactly soluble due to all-to-all scattering with a centre of mass constraint. It was initially introduced as a soluble example of a correlated metal to Mott insulator transition at half filling \cite{Hatsugai1992}. Recently, it has received renewed interest shedding light on superconductivity in  doped Mott insulators~\cite{Phillips2020,Zhu2021a,Zhu2021b,Zhao2022}. Furthermore, HK-type interactions have been used for studying interaction effects in the Haldane model \cite{Mai2022}, the Kondo effect \cite{Setty2021}, the periodic Anderson model \cite{Zhong2022}, the gapping of Weyl nodes~\cite{meng2019unpaired} or non-equilibrium physics \cite{Nesselrodt2021}. It has been argued that the metal insulator transition in the tractable HK model and the intractable Hubbard model are controlled by the same fixpoint \cite{Huang2022}. 

In this work we study the LL spectrum of the doped HK model and the resulting anomalous QOs. At finite magnetic field the solubility is only partially lost. Remarkably, the LL degeneracy is retained exactly but different LLs are strongly interacting. Hence, we can study the little explored effect of LL mixing/repulsion on LL spectra and QOs. Due to the HK interaction the effective degrees of freedom are simplified enormously. We find an exact functional form of the interaction vertex which allows for an efficient numerical treatment in the thermodynamic limit as well as further approximation to a classical Hamiltonian amenable to Monte--Carlo simulations. As a result, we find that strong LL repulsion leads to aperiodic QOs at odds with the Onsager relation. In addition, we discover unconventional temperature dependencies of QO amplitudes and effective mass renormalizations beyond LK theory. Finally, we show that the inter-LL components of the standard Hubbard interaction lead to a similar phenomenology, which highlights the general relevance of LL repulsion for interpreting QO spectra of strongly correlated quantum materials.

The paper is organized as follows. 
In sec.~\ref{sec:overview} we summarizes our main findings.
Sec.~\ref{sec:III} introduces the HK model and the continuum version for calculating the exact LL spectrum. In sec.~\ref{sec:IV} we show how to solve the model in the LL basis, discuss analytical results of the interaction vertex and use exact diagonalization and Monte--Carlo simulations to calculate QOs. In sec.~\ref{sec:Hubb} we show that the LL repulsion arising from the standard local Hubbard interaction gives rise to similar anomalous QOs as in the HK model. We discuss our findings in sec.~\ref{sec:discussion} and close with explaining the broader implications of our work in sec.~\ref{sec:conclusion}.

\section{Overview}
\label{sec:overview}
\begin{figure}
    \centering
    \includegraphics[width=\columnwidth]{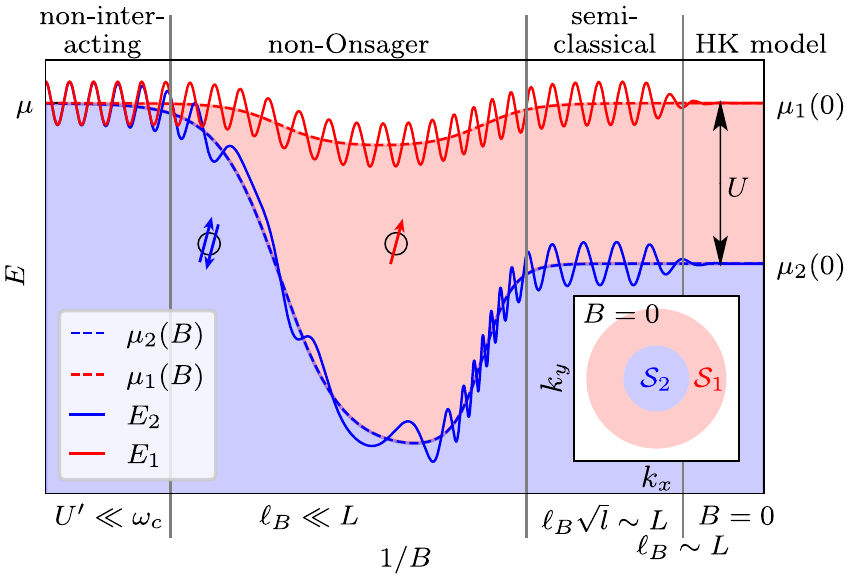}
    \caption{Schematic image of the DOS and QO of the singly and doubly occupied GS energies $E_1$ and $E_2$. In the HK model at $B=0$ momentum states are double occupied up to $\mu_2(0)=\mu-U$ and single occupied from $\mu_2(0)$ to $\mu_1(0)=\mu$ where $\mu$ is the Fermi energy in the non-interacting limit, see inset and right part. In the main panel we plot the entire energetic region where the LLs are double (single, not) occupied in blue (red, white), neglecting the LL substructure. The effective pseudo Fermi energies $\mu_i(B)$ (dashed) depend on the magnetic field and lead to QOs of the GS energy $E_1+E_2$ whose frequencies are set by $\mu_i(B)$. Different regimes emerge for increasing magnetic field going from right to left: In the semiclassical regime for sufficiently low $B$ two QO frequencies can be observed, each associated with the pseudo Fermi seas $\mac{S}_i$ at $B=0$. For higher magnetic fields the semiclassical behavior breaks down: The LLs interact and transitions between them are allowed. This interaction leads to a $B$-field dependence of the effective pseudo Fermi energies $\mu_i(B)$ which set the QO frequencies. The QOs become aperiodic. For high magnetic fields the LLs are strongly localized at odds with the center of mass constraint, such that the effective interaction $U'$ reduces to 0.}
    \label{fig:fig1}
\end{figure}
The HK model is an exactly solvable Hubbard-like model in which integrability is achieved by an infinite ranged interaction \cite{Hatsugai1992} leading to a block-diagonalized Hamiltonian 
\begin{align}
H =& \sum_{\v{k}} \epsilon_{\v{k}} (n_{\v{k},\ua} + n_{\v{k}, \da}) + U n_{\v{k}, \ua} n_{\v{k}, \da}.
\label{eq:H_HK_kspace}
\end{align}
At each momentum, the local Hilbert space is 4-dimensional consisting of the states $\ket{0_{\v{k}}},\ket{\ua_{\v{k}}},\ket{\da_{\v{k}}}$ and $\ket{\ua\da_{\v{k}}}$ with energies $0, \epsilon_{\v{k}}$ and $2 \epsilon_{\v{k}}+U$. One can then minimize the energy for each momentum and the ground state (GS) is a simple product state thereof.

The GS for any interaction strength can be understood easily from the non-interacting limit. For $U=0$ all states below the Fermi energy $\mu$ are double occupied, leading to an ordinary Fermi sea. When turning on repulsive interactions $U>0$, doubly occupied momentum states pay an energy penalty $U$. Hence, states close to the original Fermi energy avoid double occupancy giving rise to states with a single up or down electron. As a result, a single occupied region $\mac{S}_1$ forms which includes all states with energy $\mu-U<\epsilon_{\v{k}}<\mu$, whereas in the region $\mac{S}_2$ with states fulfilling $\epsilon_{\v{k}} <\mu-U$ momenta remain double occupied, see inset of Fig.~\ref{fig:fig1}. At half filling and for large repulsion $U$ a Mott insulating state emerges with a fully singly occupied band. 

In the doped Mott insulator regime the occupation regions $\mac{S}_i$ can be understood as pseudo Fermi seas. We refer to the occupation edges as pseudo Fermi surfaces (pFS) associated with the effective pseudo Fermi energies $\mu_i(0)$, where $\mu_1(0) = \mu$ and $\mu_2(0) = \mu-U$. 
%Ref.~\cite{Phillips2020} established that $\mac{S}_1$ is a Luttinger surface, because the Green's function has its zeros there.
While at first glance, the metallic regime of the HK model seems to be analogous to a two-band metal, the interacting nature is manifest in the unconventional excitations~\cite{Phillips2020} and thermodynamic properties~\cite{Hatsugai1992} as detailed below.

Remarkably, we find that the application of an orbital magnetic field, which introduces the new length scale $\ell_B = \frac{1}{\sqrt{eB}}$, 
%to the system. From the semiclassical theory of metals it is known that a magnetic field leads to LL quantization of the electrons. Prominent effects of this energy quench to highly degenerate LLs are the fractional and IQHE. The IQHE is a non-interacting effect and is also present in the 2D HK model. The interactions only effect the critical value of the magnetic field where new edge states form but not nature of the phases themselves. In contrast, the fractional quantum Hall effect is a result of interactions breaking the LL degeneracy. In the HK model however, the interaction does not lead to fractionalized state because it does not break the LL degeneracy. 
conserves the full LL degeneracy with interesting implications. First, it simplifies the many-body problem enormously by simplifying the degrees of freedom, e.g. only the LL index of the wave functions is relevant, which offers the opportunity to study solely the effects of LL mixing/repulsion. Second, we can directly work in the thermodynamic limit which allows us to derive the interaction vertex analytically. The resulting many-body problem can be efficiently solved numerically. 

 A direct application of Onsager's semiclassical theory to the HK model would lead to two distinct QO frequencies  for each of the two pseudo Fermi surfaces (pFSs) $\mu_i$ with conventional LK behaviour~\cite{Yang2021}. As one of our main results, we show that Onsager's relation is only correct in the {\it semiclassical} regime at small magnetic fields where the size of the semiclassical orbit, i.e. the characteristic size of the LLs at the Fermi energy $\sqrt{2 l^\star} \ell_B$, with the highest occupied LL $l^\star \approx \nicefrac{\mu}{\omega_c}$, is the dominant lengthscale of the system.

The reason for the appearance of a ``semiclassical'' regime in interacting metals is very generic. For low magnetic field, i.e. large $\ell_B$, multiple LLs are occupied. Inside the region $\ell_B \sqrt{l/5}$ which can be of macroscopic size, they resemble plane waves. Hence, any interaction has the same influence on high LLs at small magnetic fields as on momentum eigenstates. Therefore, the assumptions of Onsager's and LK theory, where the properties of the oscillations can be connected to electronic properties of the metal in zero magnetic field, remain true. However, we show that even in the semiclassical regime of the  HK model QOs can have a temperature dependent frequency drift because of the non-Fermi--Dirac distribution of excitations, see sec.~\ref{sec:semiclassicalQO}.

Beyond the semiclassical regime LL repulsion becomes important. Surprisingly, we observe numerically that a simple scenario of individual LLs persists. Concretely, the ground state (GS) remains close to a state with an integer occupation of each LL, see Fig.~\ref{fig:fig3}~(b). Qualitatively similar to the $B=0$ case, a double occupied region forms at low energies and single occupied one for higher energies. However, as our main result we find that the size of the regions now depend on the magnetic field $\mu_i = \mu_i(B)$, see Fig.~\ref{fig:fig1} which leads to a breakdown of Onsager's relation with aperiodic QOs.
A detailed study of the QOs in the strongly correlated non-Onsager regime, see Fig.~\ref{fig:fig4},\ref{fig:fig5} and \ref{fig:LKplots} shows that non-trivial sum and combination frequencies appear in the QO  spectrum. Finally, while all frequencies show a LK temperature dependence, they feature unusual effective mass renormalization at odds with the canonical LK theory.

\subsection{A word of caution}
As any fine-tuned exactly soluble Hamiltonian the HK model should not be considered a microscopic description of (doped) Mott insulating materials. 
Nevertheless, it can show generic physics which needs to be separated from artificial behavior originating from the infinite-ranged interactions. Concretely, the strength of the interaction between LLs is governed by two different effects. First, the deviation of the LL wavefunctions compared to  plane waves leads to a very natural change of the repulsion between LLs with opposite spin. It reduces the double occupied region $\mac{S}_2$ for multiple occupied LLs stronger than for higher magnetic fields where less LLs are occupied. Secondly, there is an artificial reduction of the effective interaction $U' = \frac{\ell_B}{L} U$ between LLs: With increasing magnetic field LLs become more localized, eventually decreasing the possibility for centre of mass conserving scattering events and, hence, the effective interaction approaches an artificial non-interacting limit in the high field regime.

In order to discuss the effect of LL repulsion beyond the HK limit, we note that the HK interaction is essentially the $\v{q}=0$, $\v{k}=\v{k}'$ part of the standard Hubbard interaction in momentum space $\tilde U \sum_{\v{k},\v{k}',\v{q}}c^\dag_{\v{k}-\v{q},\ua} c_{\v{k},\ua} c ^\dag_{\v{k}'+\v{q},\da} c_{\v{k}',\da}$. One can then study the effect of LL repulsion by projecting the Hubbard term into the LL basis and keep only inter-LL interactions but ignore LL degeneracy lifting contributions. Remarkably, in sec.~\ref{sec:Hubb} we show that we find similar aperiodic QO beyond the Onsager and LK paradigm, see Fig.~\ref{fig:occup_Hubb}.

Overall, we argue that breaking Onsager's relation is a generic effect of strongly interacting metals  with strong LL repulsion. In practice this might occur as an additional effect on top of LL degeneracy lifting effects. Our work focuses solely on the influence of interactions on LL mixing, which can be studied in a controlled way in the HK limit. It should therefore be seen as the opposite limit to standard treatments of interactions in quantum hall physics where LL mixing is only treated perturbativley and  interactions are projected into individual LLs.

%Note that for our numerical simulations of the HK model we work with infinite system sizes such that we do not expect to see a semiclassical regime, but rather a direct drop to the point with minimal $\mu_2$.

\section{Recap of the Hatsugai–Kohmoto model}
\label{sec:III}
The HK model \cite{Hatsugai1992} is described by the Hamiltonian 
\begin{align}
H =& -t \sum_{\langle \v{r}, \v{r}' \rangle, \sigma} c_{\v{r},\sigma}^\dag c_{\v{r}', \sigma} \nonumber \\
&+ \frac{U}{L^2}\sum_{\v{r}_1,\v{r}_2,\v{r}_3,\v{r}_4} \delta_{\v{r}_1+\v{r}_3,\v{r}_2+\v{r}_4} c_{\v{r}_1,\ua}^\dag c_{\v{r}_2,\ua} c_{\v{r}_3,\da}^\dag c_{\v{r}_4,\da}
\label{eq:H_HK_realspace}
\end{align}
where $L$ is the linear length of the system. We measure all lengthscales in terms of the dimensionless lattice constant $a=1$. The interaction is of infinite range and may be interpreted as centre of mass scattering: A pair of a spin-up and down electrons is scattered to a different location but their centre of mass coordinate is conserved.

The HK model can be block-diagonalized to \eqref{eq:H_HK_kspace} by simple Fourier transformation of the creation and annihilation operators
\begin{align}
c_{\v{k}} &= \frac{1}{L} \sum_{\v{r}} \e^{-\ii \v{k} \v{r}} c_{\v{r}},
\label{eq:FourierTransform_operator1}
\end{align}
see appendix~\ref{app:A}.
%c_{\v{k}}^\dag &= \frac{1}{L} \sum_{\v{r}} \e^{\ii \v{k} \v{r}} c_{\v{r}}^\dag.
%\label{eq:FourierTransform_operator2}

Initially, Hatsugai and Kohmoto \cite{Hatsugai1992} introduced the model as a simplified yet soluble version for an interaction driven metal insulator transition at half-filling. Away from the 'Mott-insulating' half-filling limit the model is metallic. However, it is not a simple Fermi liquid but features for a non-zero interaction $U$ singly $\mac{S}_1$, doubly $\mac{S}_2$ and non-occupied $\mac{S}_0$ regions in the Brillouin zone with pFSs separating them. It is then a natural question to ask, whether these pFS give rise to QOs similar to an ordinary metal? 

The GS of the HK model is highly degenerate. Each momentum state in $\mac{S}_1$ can be either occupied by a spin-up or down electron. However, this degeneracy is artificial, i.e. it is unstable against perturbations. Projecting a local Hubbard term $\tilde{U} n_{\v{r} \ua} n_{\v{r}\da}$ into the GS manifold results in an effective ferromagnetic interaction implying that the spins of the electrons inside $\mac{S}_1$ point all in the same direction \cite{Yang2021}. Henceforth, we take 
\begin{equation}
\ket{GS_\sigma} = \prod_{\v{k}_1 \in \mac{S}_1} c^\dag_{\v{k}_1 \sigma} \prod_{\v{k}_2 \in \mac{S}_2} c^\dag_{\v{k}_2 \ua} c^\dag_{\v{k}_2 \da} \ket{0}.
\end{equation}
as the robust GS.

\begin{figure}
    \centering
    \includegraphics[width=\columnwidth]{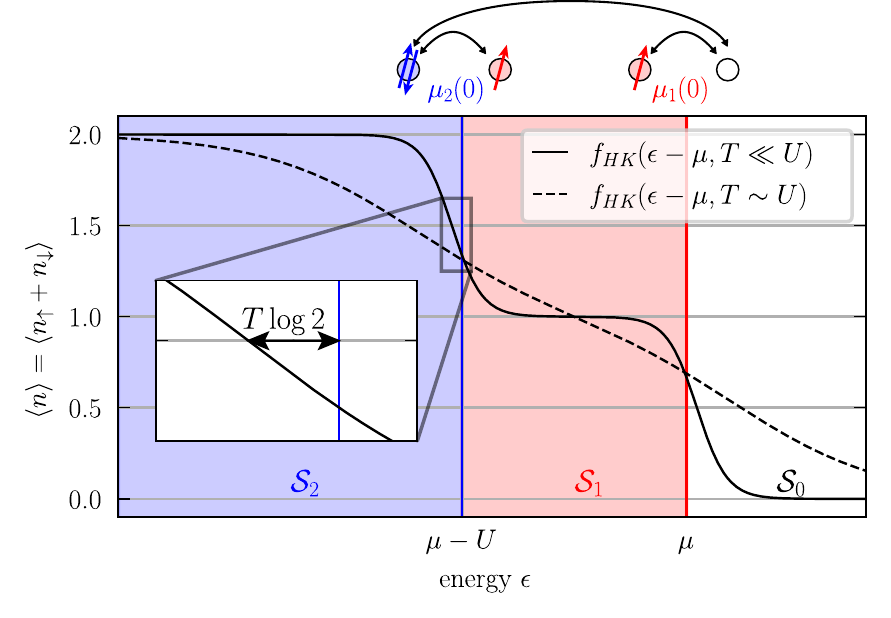}
    \caption{The HK model features at $T=0$ regions $\mac{S}_i$ in the Brillouin zone in which $\langle n_{\v{k}} \rangle = i$. $\mac{S}_2$ ($\mac{S}_1$) is bound by its pseudo Fermi energies $\mu_2$ ($\mu_1$) in blue (red). At finite temperature the occupation steps broaden asymmetrically, see zoom-in, with the  distribution function $f_{HK}$ (black, solid) due to excitations which can be excited from $\mac{S}_2$ directly to $\mac{S}_0$, see above the plot. At high temperatures $T\gtrsim U$ the details of the interaction are washed out (dashed).}
    \label{fig:fig2}
\end{figure}

All finite temperature thermodynamic properties of the HK model can be calculated exactly \cite{Hatsugai1992}. Here we only show the distribution function because it already offers a glimpse into the interacting nature of the doped Mott insulator. The partition function
\begin{align}
Z &= \tr \e^{-\beta(H-\mu N)} \\
&= \prod_{\v{k}} \left(1+2\e^{-\beta(\epsilon_{\v{k}}-\mu)}+\e^{-2\beta(\epsilon_{\v{k}}-\mu)-\beta U}\right)
\end{align}
leads to the non-Fermi--Dirac distribution function $f_{HK}(\epsilon-\mu,T)$ for the occupation number $\langle n_\ua + n_\da \rangle $ where
\begin{equation}
f_{HK}(\epsilon,T) = 2 \frac{\e^{-\beta\epsilon} + \e^{-2\beta\epsilon-\beta U}}{1+2 \e^{-\beta\epsilon} + \e^{-2\beta\epsilon-\beta U}},
\end{equation}
see Fig.~\ref{fig:fig2}. For $T\gtrsim U$ all details of the interaction are essentially washed out by temperature and the thermodynamic properties resemble those of an ordinary metal. The interesting limiting case is $T \ll U$, where
\begin{equation}
f_{HK}(\epsilon,T) \rightarrow \left[f(\epsilon+U+T\log 2,T)+1\right]f(\epsilon-T\log 2,T)
\label{eq:fHK_limit}
\end{equation}
is the combination of two Fermi--Dirac distribution functions $f$. Each occupation edge in the HK-model broadens in a Fermi--Dirac fashion with temperature, however an asymmetry of the excitations leads to a slight temperature shift of the pseudo Fermi energies, see Fig.~\ref{fig:fig2}.

Finally, note that the dispersion of the band $\epsilon_{\v{k}}$ can be of any type, depending on the form of the non-interacting part of the Hamiltonian. Throughout this manuscript we fix $\epsilon_{\v{k}} = \frac{\v{k}^2}{2m}$ corresponding to the continuous real-space term $-c^\dag(\v{r}) \frac{\v{\nabla}^2}{2m} c(\v{r})$ in order to calculate the exact LL spectrum. Formally the continuous approximation applies only for low fillings of a typical band, but we expect our findings to be generic for doped Mott insulators because the qualitative feature of two pFSs with singly and doubly occupied states persist. Note that by introducing an unbounded band structure, we loose the concept of bandwidth which is responsible for the Mott transition. This could be artificially restored by introducing a UV cutoff.

\section{Landau level interactions}
\label{sec:IV}
\subsection{Transformation to LL eigenstates}
We apply a magnetic field in $z$-direction which is perpendicular to the HK model lying in the $x$-$y$-plane, and use standard minimal coupling $-\ii \v{\nabla} \rightarrow -\ii \v{\nabla} - e \v{A}$ in the Landau gauge $\v{A} = (-B y,0,0)^\trans$. Note that the interaction does not couple to the magnetic field.

We transform to the LL basis
\begin{align}
c_{l, k_x,\sigma} &= \sum_{x,y} \Phi_{l,k_x}(x,y) c_{(x,y),\sigma} \label{eq:LLtransform_operator1}
\end{align}
%c^\dag_{l, k_x,\sigma} &= \sum_{x,y} \Phi^*_{l,k_x}(x,y) c^\dag_{(x,y),\sigma}.
%\label{eq:LLtransform_operator2}
with the LL wavefunction
\begin{equation}
\Phi_{l,k_x}(x,y) = \frac{\e^{-\ii k_x x}}{\sqrt{L \ell_B}}  \psi_l\left(\frac{y}{\ell_B}+k_x \ell_B \right)
\end{equation}
where 
\begin{equation}
\psi_l(\xi) = \frac{1}{\sqrt{2^l l! \sqrt{\pi}}} \e^{-\frac{1}{2}\xi^2} H_l\left(\xi\right)
\label{eq:psi_def}
\end{equation}
are the normalized wave functions of the quantum harmonic oscillator and $H_l$ are the (physicist's) Hermite polynomials.
The above transformation diagonalizes the non-interacting part of the Hamiltonian and gives the well known LL Hamiltonian where each LL state labeled by $l$ is \mbox{$N_\Phi=\frac{2 \pi L^2}{\ell_B^2}$-fold} degenerate. 

One of the key simplifications of the HK interaction is that the LL transformation makes it block diagonal: The interaction only couples states with different LLs $l_i$ but same momenta, giving rise to an interaction vertex $\mac{V}_{l_1l_2l_3l_4}^{L/(2\ell_B)}(k_x)$. In general, the vertex $\mac{V}_{l_1l_2l_3l_4}^{L/(2\ell_B)}(k_x)$ for a finite sized system is a difficult 3-dimensional integral which needs to be carefully solved numerically as detailed in the appendix and benchmarked in Fig.~\ref{fig:app:integrals}. Remarkably, we find that in the thermodynamic limit $L \to \infty$ all integrals of the vertex $\mac{V}_{l_1l_2l_3l_4}^{\infty}(k_x) = V_{l_1l_2l_3l_4}$ can be solved analytically, see appendix \ref{app:vertex_exact_integrals}. The full interacting Hamiltonian then reads
\begin{align}
H =& \sum_{l, k_x,\sigma} \omega_c \left(l+\frac{1}{2}\right) c^\dag_{l, k_x,\sigma}c_{l, k_x,\sigma} \nonumber\\
&+ U \frac{\ell_B}{L} \sum_{k_x,l_1,l_2,l_3,l_4} V_{l_1l_2l_3l_4} c^\dag_{l_1,k_x,\ua} c_{l_2,k_x,\ua} c^\dag_{l_3,k_x,\da} c_{l_4,k_x,\da}
\label{eq:H_HK_LL}
\end{align}
and is diagonal in $k_x$. Note that, the prefactor $\ell_B/L = \sqrt{2\pi/N_\Phi}$ normalizes the multiple sums of the interaction and hence the interaction can {\it not} be treated perturbativley in the thermodynamic limit. 
%\replaced[id=VL]{Note that the prefactor $(\ell_B/L)^{1-d_f/2} = (2\pi/N_\Phi)^{1-d_f/2}$ is responsible for the artifical non-interacting high field limit. It results from the fact that the HK interaction imposes only a single constraint $d_f=1$ on the interacting particles (whereas Coulomb interaction $d_f=2$ and Hubbard interaction $d_f=3$).}{Note, to avoid the artifical non-interacting limit we keep a nonzero prefactor $U \frac{\ell_B}{L}$ even for the infinite system vertex.} {\bf JK: Check last sentence! VL: Not correct. This is responsible for the artificial behaviour.}

We have simulated the above Hamiltonian for up to 10 LLs with exact diagonalization (ED). We emphasize that the required lattice size for a real-space calculation would be beyond any numerical capabilities. The reason why the HK model can be efficiently simulated in an orbital magnetic field has its origin in the center of mass preserving interaction which does not mix different momenta, thus, retains the full LL degeneracy. Note that this is the opposite limit of most studies of the FQHE, which usually ignore LL mixing and only treat interactions projected to individual LLs. 

\subsection{The semiclassical regime}
\label{sec:semiclassicalQO}
Before studying generic field strengths, we discuss the limit of small orbital magnetic fields. The application of a magnetic field introduces a new lengthscale, the magnetic length $\ell_B$ which may be interpreted as the size of a flux quantum $\Phi_0=(2\pi e)^{-1}$. The cyclotron orbits, i.e. characteristic size of the highest occupied LL, are much larger with a radius of $\ell_B \sqrt{2l}$ \cite{Shoenberg}. For small magnetic fields only few fluxes are inserted into the system, and the semiclassical cyclotron orbits are of macroscopic size approaching $L$. In this limit the semiclassical theory always remains valid, independent of the form of the interaction.

A quantum mechanical argument for the validity of the semiclassical theory is that inside the real space region $|y| < \ell_B \sqrt{l/5}$ LLs with index $l$ resemble plane waves
\begin{equation}
\psi_l^\infty(\xi) = \left(\frac{2}{\pi^2 l}\right)^\frac{1}{4} \cos \left(\sqrt{2l}\xi-l\frac{\pi}{2}\right),
\label{eq:psi_asymptotic}
\end{equation}
see appendix~\ref{app:B1}. For low magnetic fields, leading to LLs with a large LL index $l$ at the Fermi energy, this region is of macroscopic size. Hence, high LLs interact with exactly the same interaction as momentum states interact at zero magnetic field. Our semiclassical intuition carries over and Onsager's theorem remains valid.

The above statement applies for any metal and we now focus on the specific case of the HK model. Using the asymptotic form of the wavefunctions $\psi_l^\infty$ we evaluate the vertex  $\mac{V}_{l_1l_2l_3l_4}^{L/\ell_B}$, see appendix~\ref{app:B2}. Remarkably, we find that for sufficiently high LLs the vertex becomes diagonal in each LL leading to a `LL-HK' Hamiltonian
\begin{equation}
H_{sc} = \sum_{l, k_x} \omega_c \left(l+\frac{1}{2}\right) \left(n_{l, k_x,\ua}+n_{l, k_x,\da}\right) + U' n_{l, k_x,\ua} n_{l, k_x,\da}
\label{eq:H_HK_in_LL}
\end{equation}
which is exactly the same as in zero magnetic field, but for quantum numbers $l,k_x$. All known concepts from $B=0$ carry over exactly: LLs with $\epsilon_l < \mu-U'$ are double occupied, LLs with $\epsilon_l < \mu$ are single occupied and higher energetic LLs are not occupied. The occupation edges at $\mu$ and $\mu-U'$ lead at $T=0$ to QO with frequencies $\frac{\mu}{\omega_c}$ and $\frac{\mu-U'}{\omega_c}$ which are indeed the areas of the pFSs. 

Nevertheless the non-Fermi--Dirac distribution function of the HK model leads to unconventional behavior at non-zero temperature $T>0$. We focus on the limit $T \ll U'$ otherwise the effects of the interaction are washed out by temperature. Hence, we can make use of the approximate representation of $f_{HK}$ in terms of the Fermi--Dirac distribution function \eqref{eq:fHK_limit} and follow earlier work e.~g. Ref.~\cite{Leeb_DiffFreq}, to derive the characteristic form of the QOs of an observable $X$ (i.e. the magnetization or resistance)
\begin{align}
X  \propto& \sum_{k>0} \cos \left(2\pi k \frac{\mu+T \log 2}{\omega_c}\right) R_T(m)
\nonumber \\
&+ \cos \left(2\pi k \frac{\mu-U'-T \log 2}{\omega_c}\right) R_T(m)
\end{align}
where $R_T(m) = \frac{2\pi^2 m \ell_B^2 T}{\sinh \left(2\pi^2 m \ell_B^2 T\right)}$ is the usual LK temperature dependence. Remarkably, the only effect of the non-Fermi--Dirac distribution function in the HK model is a temperature shift of the frequencies. 

\subsection{The non-Onsager regime: Exact treatment}
We now focus on the regime $\ell_B \ll L$ such that all integration boundaries can be extended to infinity. In this limit the vertex of the LL interaction can be computed analytically with details relegated to appendix~\ref{app:vertex_exact_integrals}. Due to the degeneracy in $k_x$, we drop the momentum index $k_x$ from here on and work with completely filled LLs which corresponds to working at fixed chemical potential. We measure the filling of a LL $n_l$ in units of the LL degeneracy $N_\Phi$. 

\begin{figure}
    \centering
    \includegraphics[width=\columnwidth]{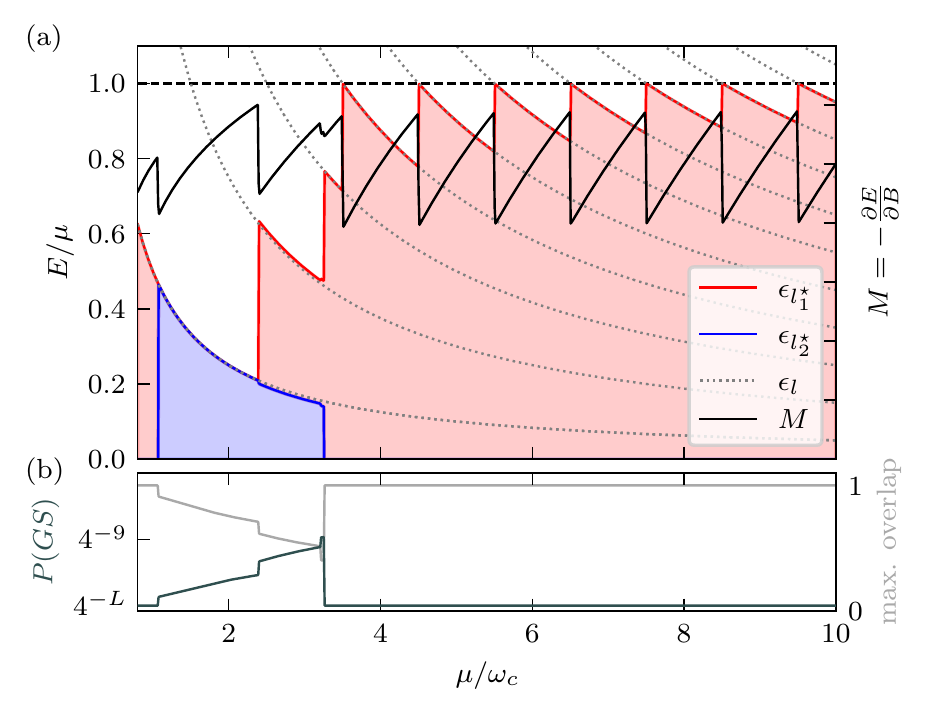}
    \caption{Panel (a): The occupation of different LLs (double occupied: transparent blue; single occupied: transparent red) is shown for inverse magnetic field $\nicefrac{\mu}{\omega_c} \propto \nicefrac{1}{B}$. The dispersion of the LLs $\epsilon_l = \omega_c \left(l+\frac{1}{2}\right)$ are shown as gray dotted lines. The red (blue) line shows the energy of the highest single (double) occupied LL $\epsilon_{l^\star_1}$ ($\epsilon_{l^\star_2}$). Jumps occur when $l^\star_1$ and $l^\star_2$ change and are also visible in the orbital magnetization (black). The data is obtained from ED with $\mathcal{L} = 10$ LLs and $\nicefrac{U'}{\mu}=\sqrt{\mu/\omega_c \mathcal{L}}$. Panel~(b) shows the many-body participation ratio $P(GS)$ of the GS on a $\log$-scale (left axis, dark gray) as well as the overlap with the closest Fock state $\max_{\alpha \in \mac{H}} \left( |\langle\alpha|GS\rangle|^2 \right)$ (right axis, light gray).}
    \label{fig:fig3}
\end{figure}

\begin{figure*}
    \centering
    \includegraphics[width=\textwidth]{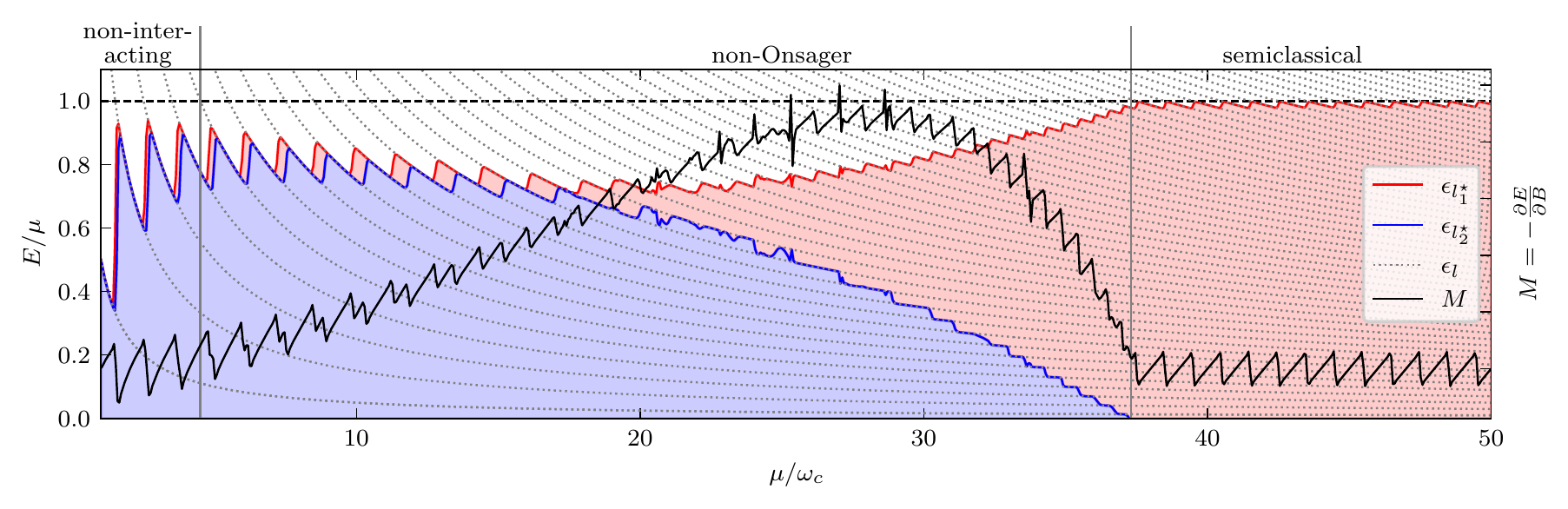}
    \caption{The occupation of different LLs (double occupied: transparent blue; single occupied: transparent red), obtained from zeroth order Monte Carlo simulations at temperatures $T \ll \omega_c$,  shown for inverse magnetic field $\nicefrac{\mu}{\omega_c} \propto \nicefrac{1}{B}$. The dispersion of the LLs $\epsilon_l = \omega_c \left(l+\frac{1}{2}\right)$ are shown as gray dotted lines. The plot should be compared to Fig.~\ref{fig:fig3} but here we simulated $\mac{L}=50$ LLs ($\nicefrac{U'}{\mu}=\sqrt{\mu/\omega_c \mathcal{L}}$). The semiclassical, non-Onsager and non-interacting regime are clearly distinguishable. The orbital magnetization $M$ experiences drops when the LL occupations change, consistent with the ED result. Additional noise in the magnetization is due to a numerical derivative of the MC data.}
    \label{fig:fig4}
\end{figure*}

Although all matrix elements of the vertex $V_{ijkl}$ can be found exactly, the resulting model remains far to complex to be solved analytically. The vertex $V_{ijkl}$ is dense and has off-diagonal and diagonal elements with no apparent substructure. Nevertheless, the transformation to the LL basis has simplified the problem enormously: First, it reduced the initial long-range interacting 2D model to a 1D long-range interacting model. Secondly, the transformation made use of the infinite system size, such that we are actually working in the thermodynamic limit and are only constrained by the number of LLs we can simulate. Overall, we can study interacting LLs with ED far beyond any real space numerical calculation.

The first remarkable result of the ED study is that even though the vertex $V_{ijkl}$ has a non-perturbative form, the exact eigenstates of the system remain close to a Fock state in the LL basis. This becomes apparent from the fact that deviations to integer filling of each LL are small, as well as from the fact that the many-body participation ratio $P^{-1}(\psi) = \dim (\mac{H}) \sum_{\alpha} \left|\braket{\alpha|\psi}\right|^4$ of the GS is small. The many-body participation ratio measures how many Fock states $\ket{\alpha}$ contribute to a many-body state $\ket{\psi}$. At the minimal value $P=(\dim(\mac{H}))^{-1}$ only a single basis state contributes, i.e. the state is a single Fock state whereas $P$ takes its maximal value of 1 for a maximally superpositioned state, e.g. $\sum_\alpha \ket{\alpha}$. 

The above results allow for a simple, perturbative understanding of the complicated vertex $V_{ijkl}$: The density-density interactions $V_{iijj}$ may be understood as ferromagnetic interactions between the LLs $i$ and $j$, because the state $c^\dag_{i \ua} c^\dag_{j \da} \ket{0}$ has a density-density interaction energy $>0$, whereas the state $c^\dag_{i \ua} c^\dag_{j \ua} \ket{0}$ has no interaction energy. Hence, the density-density interaction reduces double occupancy and aligns the electron spins of different LLs.
On the other hand the off-diagonal elements of the vertex, i.e. $i\neq j$ or $k \neq l$, stabilize antiferromagnetic LL occupations, hence also double occupancy. This contribution diagonal in LL occupation states arises as a perturbative effect via an enormous number of virtual intermediate states coupled by the off-diagonal elements of the vertex. In summary, even though the repulsive density-density interaction always wins, it is significantly reduced by the latter effect, see sec.~\ref{sec:MC}.

The second important result is that the electrons keep forming pseudo Fermi seas, i.e. energetic regions which are for LL index $l \leq l^\star_2$ doubly occupied and for LL index $l^\star_2 < l \leq l^\star_1$ singly occupied. A Fock state with these properties, which is not the exact GS but close to it, is
\begin{align}
\ket{l_1^\star,l_2^\star} = \prod_{l_1\leq l_1^\star,l_2 \leq l_2^\star} c^\dag_{l_1 \ua} c^\dag_{l_2 \da} \ket{0}.
\end{align}
We evaluate $l^\star_{1,2}$ from the exact GS by calculating
\begin{align}
l_2^\star &= \sum_l  \min \left(\left\{\langle n_{l \ua} \rangle,  \langle n_{l \da} \rangle \right\}\right)-1 \\
l_1^\star &= \sum_l  \max \left(\left\{\langle n_{l \ua} \rangle,  \langle n_{l \da} \rangle \right\}\right)-1 .
\end{align}

As stated before, it is impossible to describe the semiclassical low field regime correctly when extending the system size to infinity, which is required to evaluate the vertex analytically. However, for $U \geq \mu$ no double occupied pseudo Fermi sea $\mac{S}_2$ exists and the semiclassical regime and the low field behavior for infinite system size coincide accidentally. We focus our numerical analysis for simplicity on $U = \mu$.

\begin{figure}
    \centering
    \includegraphics[width=\columnwidth]{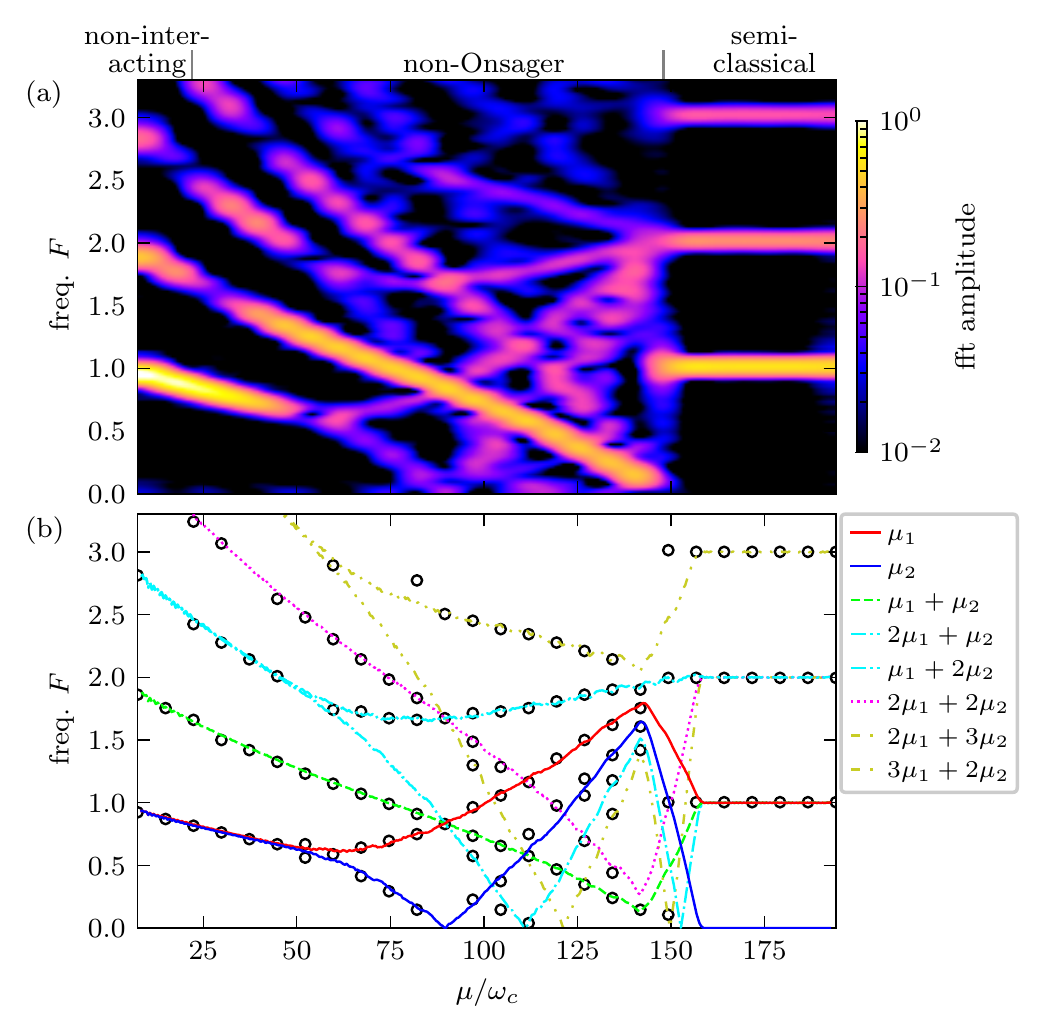}
    \caption{Panel (a) shows a STFT of the particle number $N$ for a MC data set like the one shown in Fig.~\ref{fig:fig4} but for 200 LLs at effective low temperatures. Here, we show $N$ because it is numerically more stable than $M$ which requires a derivative. However, the oscillating properties of $M$ and $N$ are the same.  Inside the semiclassical regime the Fourier spectrum shows peaks at multiples of the area of the FS. In the non-Onsager regime a plethora of peaks which are dispersive in $\nicefrac{\mu}{\omega_c}$ arise. \\
    In panel (b) we extracted the peak positions of panel~(a) (open circles). We overlayed the data points with the expected peak positions for frequencies associated with sum combinations of the effective pseudo Fermi energies $p_1 \mu_1+p_2 \mu_2$. Note that in a STFT plot frequency peaks do not appear at the actual frequencies, however the peak frequencies can be calculated from the actual frequencies, see appendix~\ref{app:stft}. Several higher orders of $(p_1,p_2)$ are visible, for clarity we focused only on the ones indicated in the legend.}
    \label{fig:fig5}
\end{figure}

\begin{figure}
    \centering
    \includegraphics[scale=1]{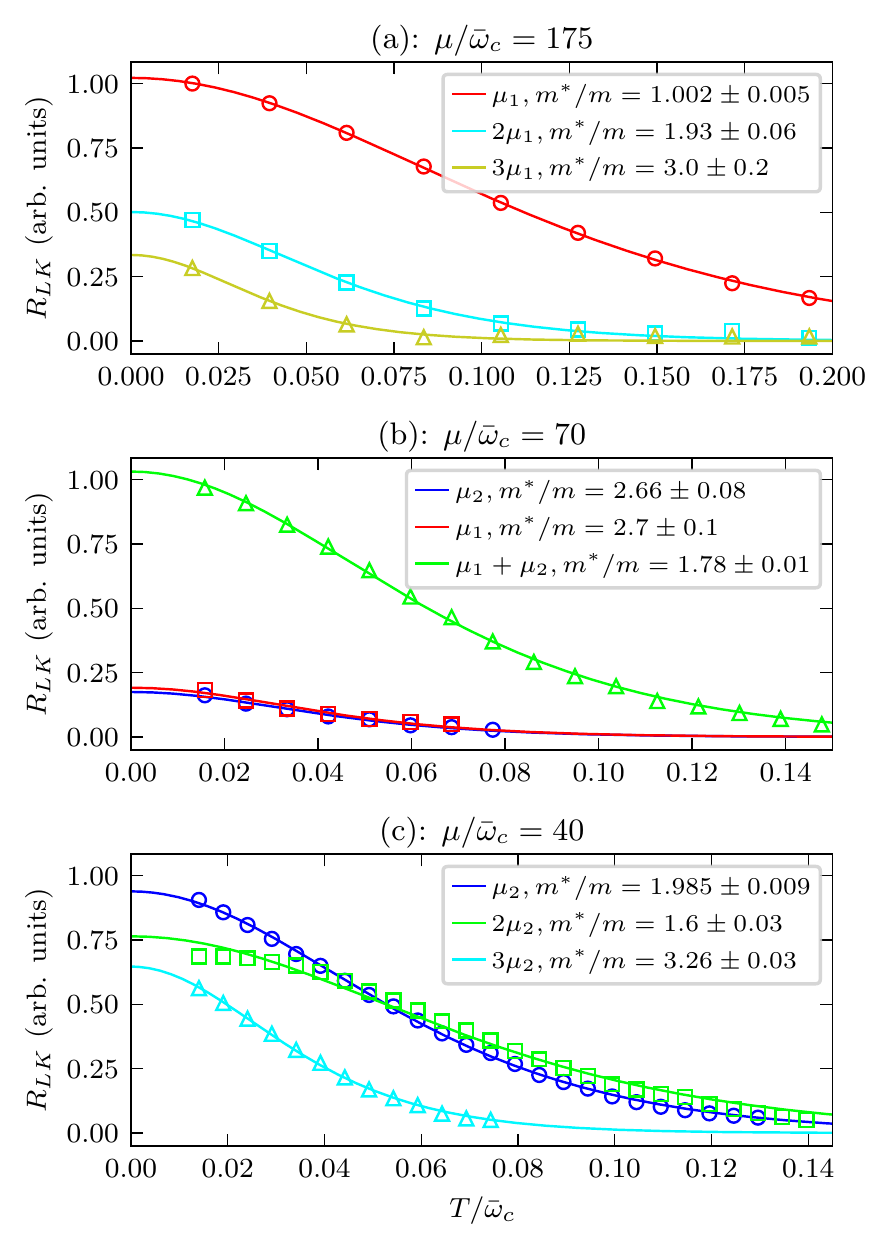}
    \caption{Temperature dependence of the main peak frequencies of Fig.~\ref{fig:fig5} (open symbols). The temperature dependence is extracted for 3 different windows, each ranging from $[\nicefrac{\mu}{\bar\omega_c}-10,\nicefrac{\mu}{\bar\omega_c}+10]$. The data is fitted with the LK factor $R_{T}(m^*)$ to obtain the effective mass (solid lines). The color coding of the frequencies is in accordance with Fig.~\ref{fig:fig5}~(b). Note that very low temperatures are not accessible due to freezing of the MC simulation.}
    \label{fig:LKplots}
\end{figure}

In Fig.~\ref{fig:fig3}~(a) the energy dispersion of the highest single (double) occupied LL $\epsilon_{l^\star_1}$ ($\epsilon_{l^\star_2}$) is shown,  as well as the orbital magnetization obtained from the GS energy as a function of $\nicefrac{\mu}{\omega_c} \propto 1/B$. For small magnetic fields, i.e. large $\nicefrac{\mu}{\omega_c}$, the number of occupied LLs drops periodically at $\nicefrac{\mu}{\omega_c} = \mathbb{Z}+\nicefrac{1}{2}$, i.e. when the energy of the highest occupied LL becomes larger than the chemical potential $\mu$. These periodic QO appear  in the magnetization in accordance with Onsager's seminal relation. However, at a sufficiently strong magnetic fields the system can minimize its energy by occupying the lowest single occupied LL with a spin down and a spin up electron, $l_2^\star$ increases by one. Similarly, it might be energetically preferable to keep the lowest double occupied LL and instead depopulate the highest single occupied LL, $l_1^\star$ decreases by one. Both processes lead to jumps in the magnetization. Importantly, these jumps are aperiodic and the critical magnetic field values where they appear depend on the details of the vertex and the interaction strength. The main conclusion is that the resulting QOs become aperiodic breaking Onsager's  relation!

Henceforth, we can understand the effect of interactions in terms of effective chemical potentials $\mu_i(B)$ for the doubly and singly occupied states, which is analogous to the $B=0$ HK model where $\mu_1(0)=\mu$ and $\mu_2(0)=\mu-U$.

\subsection{Qualitative results for many LLs: A Monte--Carlo study}
\label{sec:MC}
\subsubsection{Zero temperature}
The simple results from the ED simulations suggest that a perturbative picture where LLs remain the exact eigenstates might be sufficient to understand the underlying physics. In this picture the off-diagonal matrix elements, i.e. $V_{ijkl}$ for $i\neq j$ and $k\neq l$ are treated as perturbations to the classical Hamiltonian
\begin{equation}
H_0 = \sum_{l,\sigma} \omega_c \left(l+\frac{1}{2}\right) n_{l,\sigma} + U' \sum_{l,l'} V_{lll'l'} n_{l,\ua} n_{l',\da}.
\label{eq:H_LL_0}
\end{equation}
The eigenstates of $H_0$ are known exactly, since $[n_{l,\sigma},H_0]=0$. These are the Fock states in the LL basis $\ket{n_{0,\ua},n_{0,\da};n_{1,\ua},...}$. In principle, the energy of each eigenstate can be computed efficiently, however finding the GS by a direct calculation of all eigenstates is numerically costly. In the following, we show that an efficient way to find the GS and obtain the finite temperature dynamics with respect to $H_0$ is the use of Monte-Carlo (MC) sampling employing the Metropolis algorithm.

In principle it is possible to include perturbations of second or higher order (the first order vanishes) but in practice the dense form of the off-diagonal vertex requires to sum over a large fraction of states of the entire Hilbert space such that the second order correction of the eigenstate energy cannot be computed efficiently. By a careful comparison between ED and MC results, we have shown that even in the presence of off-diagonal interactions states remain close to LL Fock states. Thus, we can conclude that the zeroth order approximation is sufficient for a correct qualitative picture. Higher order perturbations will decrease the strength of the diagonal elements $V_{lll'l'}$ and, therefore, we observe that the zeroth order MC simulation overestimates the strength of the interaction~$U$.

Fig.~\ref{fig:fig4} shows the results of the MC simulation of \eqref{eq:H_LL_0} in the same style as Fig.~\ref{fig:fig3} for ED. The MC simulation allows to access many more LLs and hence more oscillations. We have subsequently decreased the temperature to obtain the GS occupation. 

Importantly, the MC simulations provide further numerical evidence for the schematic image sketched in Fig.~\ref{fig:fig1}: The number of doubly and singly occupied LLs set the QO frequencies. For Fig.~\ref{fig:fig5} we have collected data of 200 LLs at effectively zero temperature. Due to the fact that the QO frequencies depend on the magnetic field, we perform a short-time Fourier transformation (STFT) as $\mu/\omega_c$ changes. In the STFT small, consecutive windows of the complete data are Fourier transformed, allowing to study the magnetic field dependence of the peak frequencies (for details of the STFT method see appendix~\ref{app:stft}).

Strikingly, Fig.~\ref{fig:fig5} shows that the observed frequencies match with the effective pseudo Fermi energies $\mu_1(B) = \bar \epsilon_{l_1^\star+\nicefrac{1}{2}}$ ($\mu_2(B) = \bar\epsilon_{l_2^\star+\nicefrac{1}{2}}$) of the singly (doubly) occupied LLs, see the red (blue) solid line in Fig.~\ref{fig:fig5}~(b). Note that in a STFT the frequencies $F\left(\nicefrac{\mu}{\omega_c}\right)$ are not observed directly but due to the consecutive Fourier transformations only $\overline{F(t)} + t \overline{\frac{\d F}{\d t} (t)}$ where $\overline{\cdot(t)}$ denotes the average over the window with midpoint $t$, see appendix~\ref{app:stft}.

The most prominent feature in Fig.~\ref{fig:fig5}~(a) are not the basis frequencies but the combination frequencies $p_1 \epsilon_{l_1^\star+1/2}+p_2 \epsilon_{l_2^\star+1/2}$ with integers $p_1, p_2$. Our two main observations are: (i) In the canonical theory of QOs only multiples of the basis frequencies are allowed, i.e. $p_1 \epsilon_{l_1^\star+1/2}$ and $p_2 \epsilon_{l_2^\star+1/2}$, whereas we observe sum combinations of these basis frequencies which is highly unusual. (ii) The higher orders come with anomalous amplitudes. The sum frequency is clearly dominant in the non-Onsager regime but the canonical higher orders $(p_1,0)$ and $(0,p_2)$ with $p_1,p_2 > 1$ are absent.

The observed QOs in Fig.~\ref{fig:fig5} show a clear breakdown of Onsager's relation which would predict frequencies set by $\mu_i(0)$ with $i=1,2$ and higher harmonics thereof. Nevertheless, oscillations remain visible and they are set by the effective pseudo Fermi energies $\epsilon_{l_{1,2}^\star+1/2}$ which is determined from the interaction. The oscillatory part of a thermodynamic quantity $X_\text{osc}$ reads 
\begin{equation}
X_\text{osc} \propto \sum_{p_1,p_2 >0} A_{(p_1,p_2)} \cos \left(2\pi\frac{p_1 \epsilon_{l_1^\star+1/2}+p_2 \epsilon_{l_2^\star+1/2}}{\omega_c}\right)
\end{equation}
and some amplitudes $A_{(p_1,p_2)}$ are too small to be observed in our numerics.

\subsubsection{Finite temperature}
A further advantage of the MC simulation is that it allows for an efficient computation of finite temperature properties. Fig.~\ref{fig:LKplots} shows the temperature dependence of the amplitudes of the strongest peaks of Fig.~\ref{fig:fig5} for different windows centered around $\nicefrac{\mu}{\bar\omega_c}$. We chose windows in the semiclassical (Fig.~\ref{fig:LKplots}~(a)) as well as in the non-Onsager regime (Fig.~\ref{fig:LKplots}~(b) and (c)). Strikingly, we find for all frequencies and all windows a clear LK dependence of the amplitudes which can be traced back to the underlying Fermi--Dirac-like distribution of excitation energies. However, fitting the amplitudes with the LK factor $R_T\left(m^*\right)$ to obtain the effective mass $m^*$ of each frequency shows a breakdown of the LK theory. In the semiclassical regime the higher harmonics are damped with an effective mass being integer multiples of the bare charge carrier mass $m$, as expected, see Fig.~\ref{fig:LKplots}~(a). Contrarily, in the non-Onsager regime the sum frequency $\mu_1+\mu_2$ has the lowest effective mass $m^* \approx 1.7 m$ whereas the basis frequencies decay faster in temperature with $m^* \approx 2 \text{ to }3 m$.

Note that we do no find a clear indication for temperature drifts of the frequencies as in the semiclassical regime.

\section{Landau level repulsion in the Hubbard model}
\label{sec:Hubb}
\begin{figure}
    \centering
    \includegraphics[width=\columnwidth]{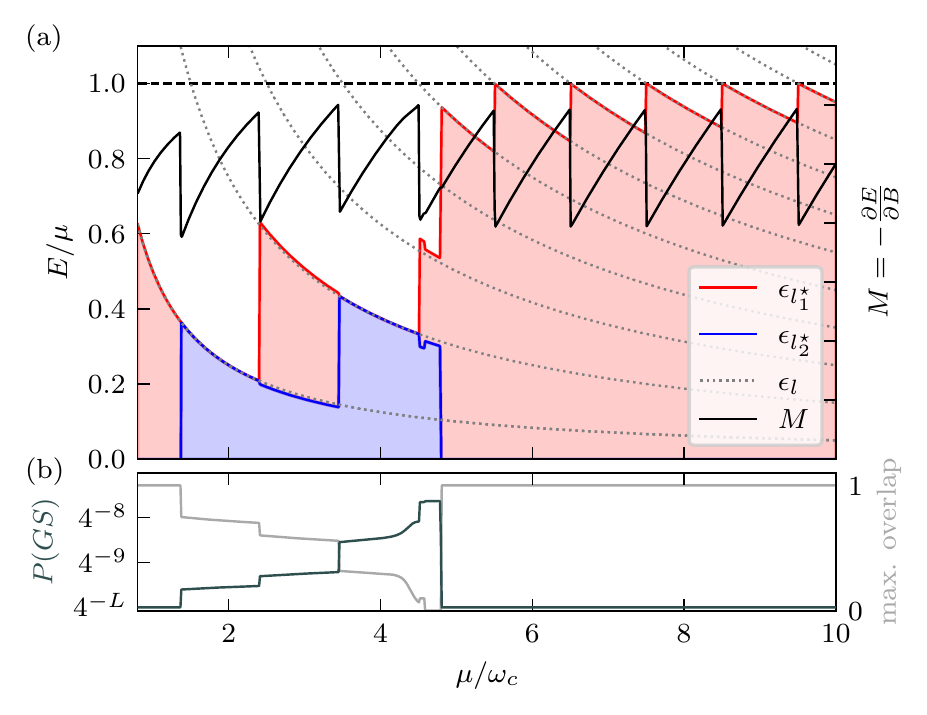}
    \caption{Panel (a): The occupation of different LLs in the Hubbard model (double occupied: transparent blue; single occupied: transparent red) is shown for inverse magnetic field $\nicefrac{\mu}{\omega_c} \propto \nicefrac{1}{B}$. The dispersion of the LLs $\epsilon_l = \omega_c \left(l+\frac{1}{2}\right)$ are shown as gray dotted lines. The red (blue) line shows the energy of the highest single (double) occupied LL $\epsilon_{l^\star_1}$ ($\epsilon_{l^\star_2}$). Jumps occur when $l^\star_1$ and $l^\star_2$ change and are also visible in the orbital magnetization (black). The data is obtained from ED with 10 LLs and $\nicefrac{\tilde U '}{\mu}=5/\sqrt{\mu/\omega_c}$. Panel~(b) shows the many-body participation ratio $P(GS)$ of the GS on a $\log$-scale (left axis, dark gray) as well as the overlap with the closest Fock state $\max_{\alpha \in \mac{H}} \left( |\langle\alpha|GS\rangle|^2 \right)$ (right axis, light gray).}
    \label{fig:occup_Hubb}
\end{figure}
The HK model provides a good starting point to explore the LL spectrum of interacting metals because its physics at zero magnetic field is well understood due to its exact solubility. However, we argue that our findings about LL repulsion leading to anomalous QOs are generic and not a pure artifact of the infinitely ranged HK interaction. In this subsection we show that we obtain similar results for the Hubbard model as summarized in Fig.~\ref{fig:occup_Hubb}.

We project the standard Hubbard interaction $\tilde U \sum_{\v{r}} n_{\v{r},\ua}n_{\v{r},\da}$ in the LL basis and ignore contributions lifting the LL degeneracy, e.g. its $k_x$ momentum dependence. Analogously to the HK model we obtain 
\begin{align}
\tilde H =& \sum_{l, k_x,\sigma} \omega_c \left(l+\frac{1}{2}\right) c^\dag_{l, k_x,\sigma}c_{l, k_x,\sigma} \nonumber\\
&+ \tilde U \frac{1}{\ell_B} \sum_{k_x,l_1,l_2,l_3,l_4} \tilde V_{l_1l_2l_3l_4} c^\dag_{l_1,k_x,\ua} c_{l_2,k_x,\ua} c^\dag_{l_3,k_x,\da} c_{l_4,k_x,\da}
\label{eq:H_Hubb_LL}
\end{align}
where Hubbard quantities are marked by a tilde. The LL vertex  $\tilde V_{ijkl}$ for the Hubbard model can also be computed exactly, see appendix~\ref{app:Hubb}, and is similarly dense and unstructured. The main difference to the HK model is that the effective interaction $\tilde U/\ell_B$ increases for high magnetic fields, causing the artifical non-interacting regime of the HK model to disappear.

We have solved \eqref{eq:H_Hubb_LL} for up to 10 LLs by ED, see Fig.~\ref{fig:occup_Hubb}. Remarkably, the results for the Hubbard model closely resemble the results of the HK model, Fig.~\ref{fig:fig3}. Concretely, LLs with $B$-field dependent effective pseudo Fermi energies remain a good description of the system leading to a breakdown of the Onsager relation for QOs.

\section{Discussion}
\label{sec:discussion}
Our approach to study QOs in a doped Mott insulator is based on the HK interaction and an exact transformation to the LL basis in the thermodynamic limit. We showed that the resulting LL vertex retains the LL degeneracy even for strong interactions. However, the interactions lead to magnetic field dependent pseudo Fermi energies due to strong repulsion between different LLs. As a result, we find QOs beyond Onsager's relation with unusual properties. The aperiodic QOs can be mainly understood on the basis of the magnetic field dependence of the pseudo Fermi energies with three notable exceptions: (i) The emergence of new QO frequencies which are the sum of pFS $\mu_1$ and $\mu_2$. (ii) The anomalous amplitudes of the different harmonics, e.g. the sum frequencies are strong whereas ordinary second or higher harmonics are absent. (iii) The unusual effective masses extracted from the LK temperature dependence of the different harmonics. 

For canonical QOs different mechanisms are known which could possibly explain the emergence of sum frequencies. However, most of them are due to processes in experimental setups, like magnetic interactions~\cite{Shoenberg}, and can therefore be ruled out. Neither can oscillations of the effective Fermi energies be the reason for observation (i), since they would lead to oscillation with sum and difference frequencies. We suggest that in strongly interacting systems the sum frequencies can be understood as oscillations of the quasiparticle lifetime \cite{Leeb_DiffFreq}. In Ref.~\cite{Leeb_DiffFreq} interband scattering by impurities leads to a coupling of LLs from different bands which gives rise to QOs of the quasiparticle lifetime. New combination frequencies of QOs appear in transport properties but no difference (only sum) frequencies are observed in thermodynamic quantities similar to the magnetization studied here. The underlying mechanism in our case is qualitatively similar, e.g. the interaction driven feedback of the different QO periods of the two occupation edges leads to sum combination frequencies in thermodynamic quantities. Consequently, we expect both sum {\it and} difference frequencies $p_1 \mu_1 + p_2 \mu_2$ with $p_1,p_2 \in \mathbb{Z}$ to appear in transport properties.

Observation (ii) and (iii) are beyond standard perturbative effects of QOs in interacting system~\cite{Shoenberg,wasserman1996influence,Allocca2021}. Especially the small effective mass of the sum frequency is in stark contrast with known theories of QOs, where sum combinations $\mu_1+\mu_2$ have temperature dependencies $R_T(m_1^*+m_2^*)$ or $R_T(m_1^*) R_T(m_2^*)$ and, hence, necessarily decay faster in temperature then their basis frequencies.

So far we have concentrated on the effect of LL repulsion on QOs but it is interesting to speculate about other non-perturbative effects. For example, LL repulsion can also lead to an interesting interplay between the IQHE effect and Mott physics. The Mott insulating state of the HK model without a magnetic field appears at half filling and $U$ being larger than the bandwidth $\Lambda$ such that the entire Brillouin zone is singly occupied. In our continuum model this can be artificially realized by introducing a UV cut-off $\Lambda = \mu_1(0)$. Applying a magnetic field leads to the formation of double occupied LLs at $B_{c0}$ by deoccupying the ``highest'' LL, analogous to Fig.~\ref{fig:fig3}. Then QOs, IQHE or FQHE would only be visible for $\mu_1(B) \neq \mu_1(0) = \Lambda$. The transition between regimes with singly and doubly occupied LLs would be accompanied by a reformation of edge states, one associated to the particle pocket at the lower Hubbard band and the other one associated to the hole pocket at the upper Hubbard band. As a result, a magnetic field induced transition between a Mott insulator and a Hall insulating state should occur with a distinct Hall response.

\section{Conclusion}
\label{sec:conclusion}
We have studied the LL spectrum of the exactly soluble HK model and the resulting QOs. The HK interaction does not break the LL degeneracy but leads to a strong repulsion between LLs. We found various exact results for the interaction vertex between LLs which allowed the efficient numerical simulation of up to ten LLs. Subsequently, we showed that the main qualitative effects can already be understood from density-density interactions between LLs, which allowed us to perform Monte Carlo simulations for hundreds of LLs. The most important effect is the emergence of effective pseudo Fermi energies $\mu_i(B)$ which depend on the magnetic field strength via the interaction vertex. 

The implications of the magnetic field dependent LL repulsion are manifold: The resulting QOs and the critical magnetic fields of IQHE transitions become aperiodic. Hence, QOs are not connected to the area of the pseudo Fermi energies at zero field in contrast to Onsager's seminal relation. Furthermore, LL interactions give rise to novel sum combination frequencies and LK temperature decays of the QO amplitudes with unusual effective mass renormalizations. 

In the future it will be interesting to explore other physical observables of the (partially) soluble HK model in an orbital magnetic field. In addition, the fine-tuned limit of infinite ranged interactions could be used as a starting point for including generic perturbations, e.g. those lifting the LL degeneracy. It would be very worthwhile to look for our aperodic QOs with numerical methods, e.g. recent extensions of DMFT to include orbital magnetic fields~\cite{vuvcivcevic2021electrical}. Similarly, other exactly soluble models~\cite{chen2018exactly} could shed light on interaction effects and QOs in doped Mott insulators and non-perturbative parton descriptions can help to map out the possible phenomenologies~\cite{he2022electronic}.   

The canonical Onsager and LK theory of QOs, which is essentially a semiclassical theory of non-interacting electrons, has been unreasonably successful. Over the last decades, it has been applied beyond its regime of validity to understand QO experiments of weakly as well as strongly correlated systems. In that context our work rationalizes that even in the strongly interacting HK model we recover canonical QOs in the semiclassical limit. However, there are by now several experimental examples of strongly correlated materials showing QOs beyond the canonical description~\cite{tan2015unconventional,hartstein2018fermi,liu2018fermi,xiang2018quantum,pezzini2018unconventional,li2020emergent,leeb2021anomalous,czajka2021oscillations}. Our study indeed provides rigorous calculations for novel aperiodic QOs with unusual mass renormalizations, and we hope it can serve as  a stepping stone for exploring new theoretical scenarios and generalizations of Onsager's relation.  

\begin{acknowledgments}
We acknowledge helpful discussions with Inti Sodemann. V.L. acknowledges support from the Studienstiftung des deutschen Volkes. J.K. acknowledges support from the Imperial- TUM flagship partnership, as well as the Munich Quantum Valley, which is supported by the Bavarian state government with funds from the Hightech Agenda Bayern Plus.
\end{acknowledgments}

\section*{Data availability}
Code and data related to this paper are available on Zenodo \cite{code} from the corresponding authors
upon reasonable request.

\clearpage

\appendix
\section{Transformation to the Landau level basis}
\label{app:A}
Here we show how the HK model becomes block diagonal by Fourier transformation, i.e. deriving \eqref{eq:H_HK_kspace} from \eqref{eq:H_HK_realspace}, and how the LL vertex arises, i.e. the derivation of \eqref{eq:H_HK_LL}.

\begin{figure*}
    \centering
    \begin{tabular}{ccc}
        \includegraphics[width=0.3\textwidth]{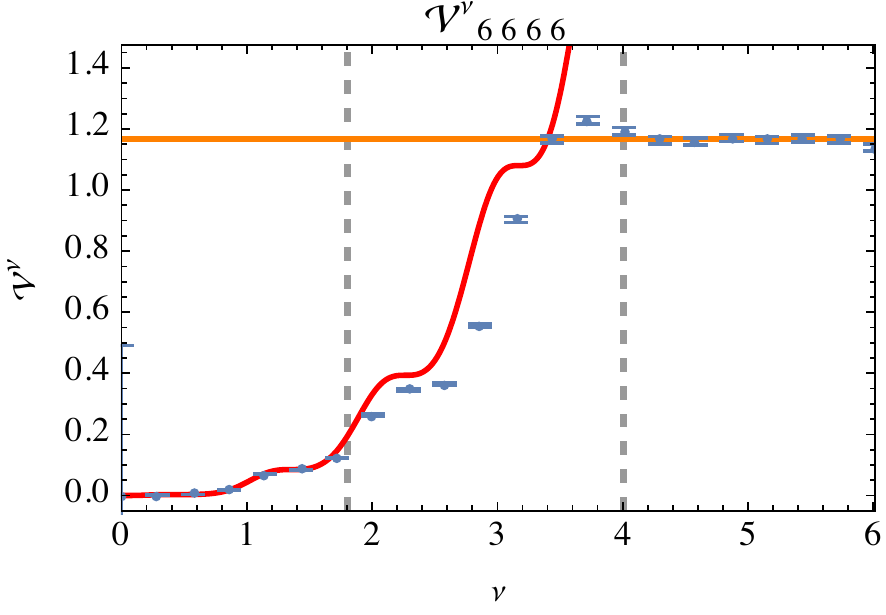} \hspace*{0.02\textwidth}&
        \includegraphics[width=0.3\textwidth]{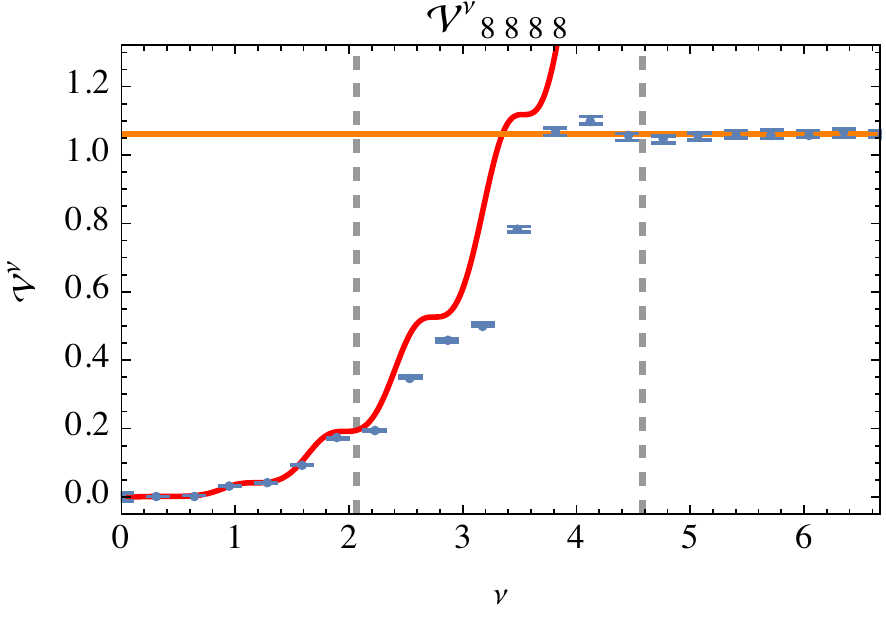} \hspace*{0.02\textwidth}&
        \includegraphics[width=0.3\textwidth]{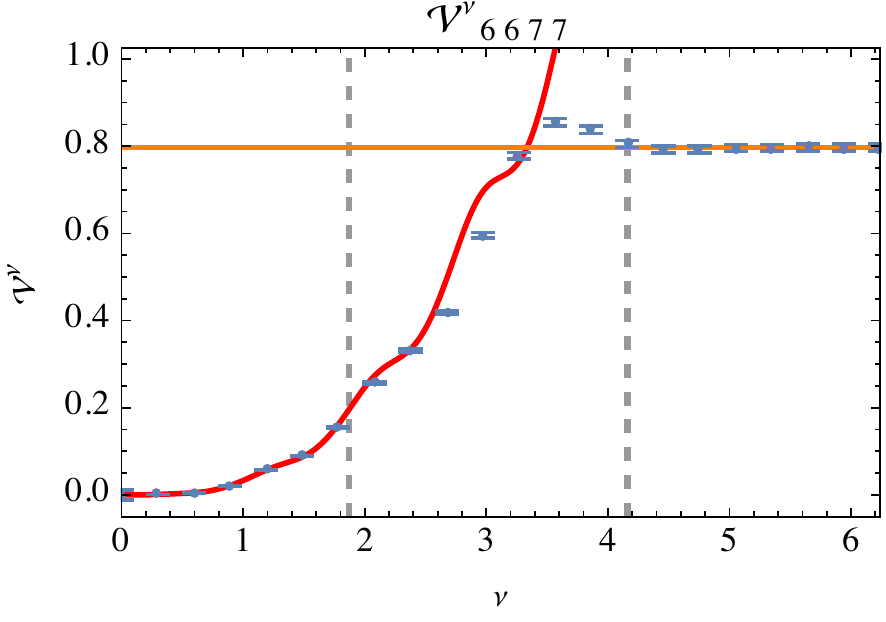}
        \vspace*{0.3cm}\\
        \includegraphics[width=0.3\textwidth]{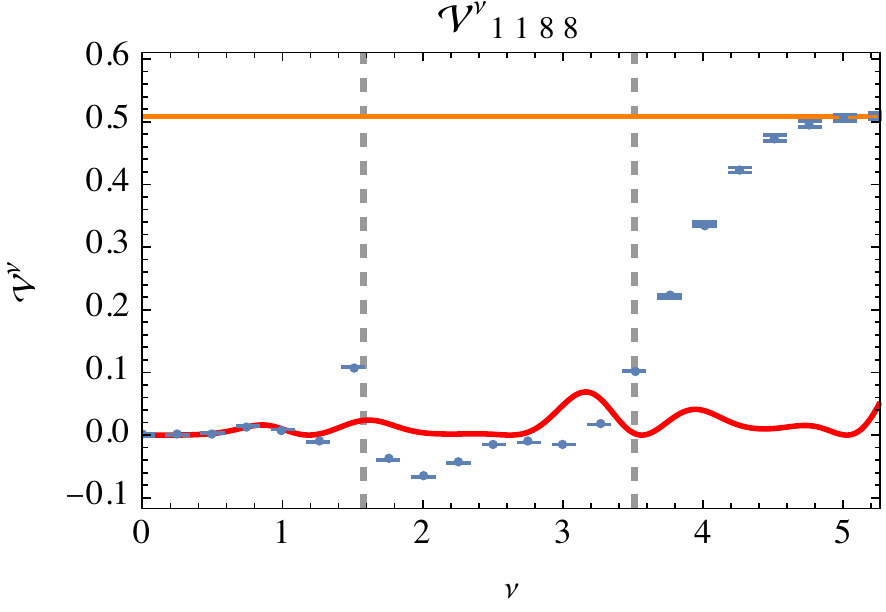} \hspace*{0.02\textwidth}&  \includegraphics[width=0.3\textwidth]{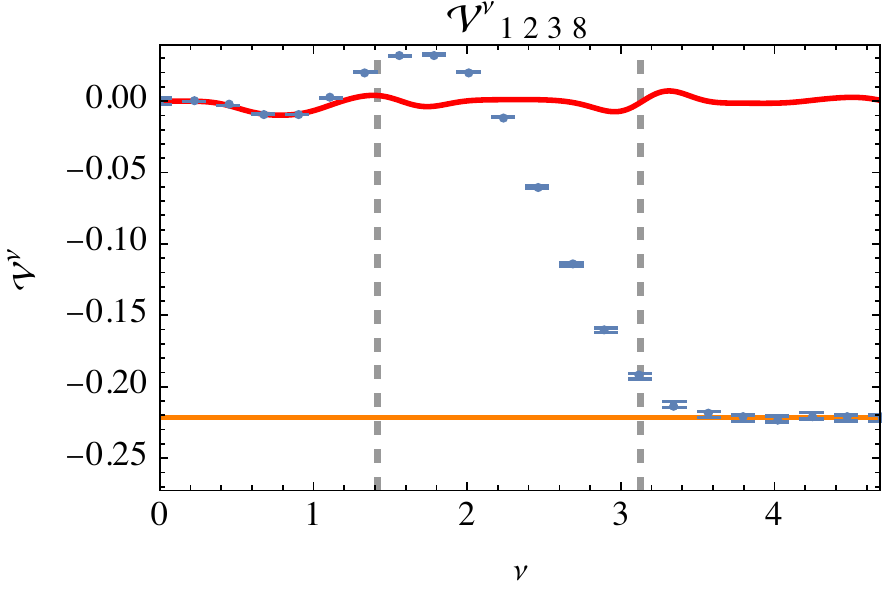}
        \hspace*{0.02\textwidth}&
        \includegraphics[width=0.3\textwidth]{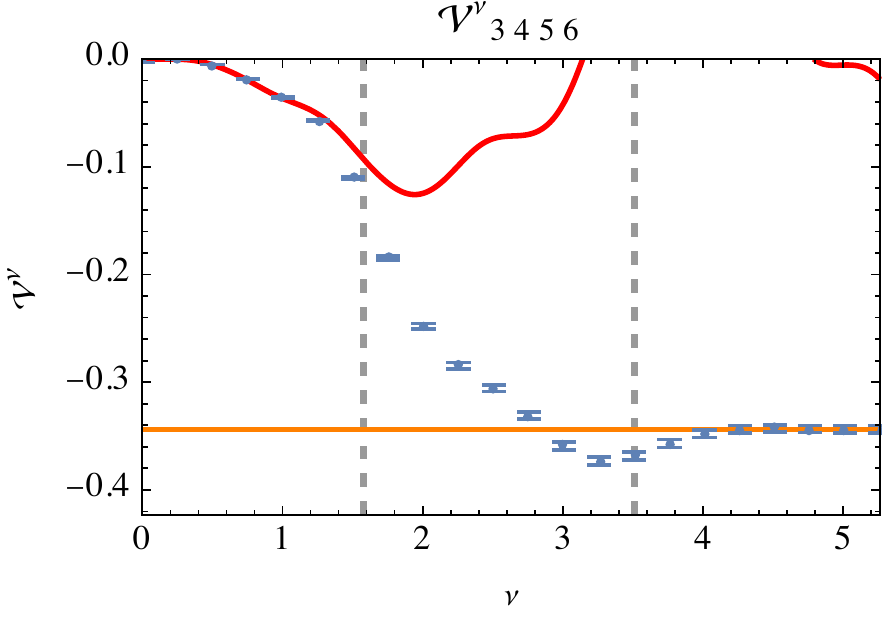}
    \end{tabular}
    
    \caption{Dependence of the integral $\mac{V}^\nu_{ijkl}(k_x=0)$ on the integration boundary $\nu$ for various indices $i,j,k,l$ (blue dots with error bars). The numerical integration is done with the built-in ``NIntegrate'' function of {\it Mathematica} 12 which returns the estimated error of the integration. For $\nu \lesssim \nicefrac{1}{2}\sqrt{2 \bar l+1}$ (gray, dashed) where $\bar l$ is the mean of the 4 indices, the vertex can be described by the semiclassical vertex \eqref{eq:app:vertex_semiclassical} (red line). By extending the integration boundaries to infinity the integral can be solved exactly, i.e. $V_{ijkl} = \mac{V}^\infty_{ijkl}$ (orange line, \eqref{eq:app_Vijkl}). This becomes approximately a good approximation when $\nu \gtrsim \xi_0(\bar l)$ (gray, dashed), where $\xi_0(l) \approx 2 \sqrt{l}$ is the value above which the LL wavefunction is exponentially small, i.e. $\xi_{0}(l) = \min \{\xi \in \mathbb{R}\text{: } \forall x >\xi \quad |\psi_l(x)| \leq 0.05\}$.}
    \label{fig:app:integrals}
\end{figure*}

We start from the real space Hamiltonian \eqref{eq:H_HK_realspace} and transform its interaction to the LL basis. For simplicity we carry out the calculation separately for the $x$ and $y$-component. We begin with the $x$-component which is for our gauge choice of the magnetic field analogous to the HK model at zero magnetic field
\begin{align}
&\frac{1}{L} \sum_{x_1,x_2,x_3,x_4} \delta_{x_1+x_3,x_2+x_4} c^\dag_{x_1,\ua} c_{x_2,\ua} c^\dag_{x_3,\da} c_{x_4,\da}
\nonumber\\
=&\frac{1}{L^3}\sum_{k_1,k_2,k_3,k_4} c^\dag_{k_1,\ua} c_{k_2,\ua} c^\dag_{k_3,\da} c_{k_4,\da}  \nonumber\\
&\times \underbrace{\sum_{x_1}\e^{\ii x_1(k_1-k_4)}}_{L\delta_{k_1,k_4}} \underbrace{\sum_{x_2} \e^{-\ii x_2(k_2-k_4)}}_{L\delta_{k_2,k_4}} \underbrace{\sum_{x_3} \e^{\ii x_3(k_3-k_4)}}_{L\delta_{k_3,k_4}}
\nonumber\\
=& \sum_{k_4} c^\dag_{k_4,\ua} c_{k_4,\ua} c^\dag_{k_4,\da} c_{k_4,\da}
\end{align}
and for the $y$-component at a given momentum $k_x$
\begin{align}
&\frac{1}{L} \sum_{y_1,y_2,y_3,y_4} \delta_{y_1+y_3,y_2+y_4} c^\dag_{y_1,\ua} c_{y_2,\ua} c^\dag_{y_3,\da} c_{y_4,\da}
\nonumber\\
=&\frac{1}{L \ell_B^2}\sum_{l_1,l_2,l_3,l_4} c^\dag_{l_1,\ua} c_{l_2,\ua} c^\dag_{l_3,\da} c_{l_4,\da} \int_{-L/2}^{L/2} \d y_1\d y_2\d y_3 \nonumber\\
& \times  \psi_{l_1}\left(\frac{y_1}{\ell_B}+\ell_B k_x\right)
\psi_{l_2}\left(\frac{y_2}{\ell_B}+\ell_B k_x\right)  \nonumber\\
& \times \psi_{l_3}\left(\frac{y_3}{\ell_B}+\ell_B k_x\right) \psi_{l_4}\left(\frac{y_1-y_2+y_3}{\ell_B}+\ell_B k_x\right)
\nonumber\\
=& \frac{\ell_B}{L}\sum_{l_1,l_2,l_3,l_4} \mac{V}_{l_1l_2l_3l_4}^{L/(2\ell_B)}(k_x) c^\dag_{l_1,\ua} c_{l_2,\ua} c^\dag_{l_3,\da} c_{l_4,\da} 
\label{eq:app:HK_derive_vertex}
\end{align}
where the general vertex is
\begin{align}
\mac{V}_{l_1l_2l_3l_4}^\nu(q) =&  \int_{-\nu}^{\nu} \d\xi_1\d\xi_2\d\xi_3 \psi_{l_1}\left(\xi_1+\ell_B q\right) \nonumber\\
& \times  
\psi_{l_2}\left(\xi_2+\ell_B q\right) \psi_{l_3}\left(\xi_3+\ell_B q\right) \nonumber\\
& \times \psi_{l_4}\left(\xi_1+\xi_3-\xi_2+\ell_B q\right).
\end{align}
Different matrix elements of the general vertex for $q=0$ are shown in Fig.~\ref{fig:app:integrals}.

\section{Semiclassical limit}
This section includes details of calculations in the semiclassical limit. We derive the asymptotic wavefunction for high LLs and derive the semiclassical vertex which is diagonal in the LL index.
\vspace{0.5cm}
\subsection{Asymptotic wavefunction for high LLs}
\label{app:B1}
In this subsection we derive \eqref{eq:psi_asymptotic} from its definition \eqref{eq:psi_def} by making use of the asymptotic form of the Hermite polynomials $H_l(x)$ \cite{Dominici2007_asymptotic} inside the region \mbox{$|\xi|<\sqrt{2l}$}
% \begin{equation}
% H_l\left(\sqrt{2l}\sin\theta\right) \approx \sqrt{\frac{2}{\cos \theta}} \e^{\frac{l}{2}\left(\log (2l) -\cos (2\theta)\right)} \cos\left(\frac{l}{2}\sin(2\theta)+l\theta+ \frac{\theta}{2} -l\frac{\pi}{2}\right)
% \end{equation}
\begin{align}
&H_l\left(\xi\right) \approx \sqrt{\frac{2}{\sqrt{1-\frac{\xi^2}{2l}}}} \e^{\frac{l}{2}\left(\log (2l) -1+\frac{\xi^2}{2l}\right)}
\nonumber\\
&\times \cos\left(\sqrt{\frac{l}{2}}\xi\sqrt{1-\frac{\xi^2}{2l}}+\left(l+\frac{1}{2}\right)\arcsin \left(\frac{\xi}{\sqrt{2l}}\right) -l\frac{\pi}{2}\right).
\end{align}
We expand the asymptotic form in $\nicefrac{\xi^2}{l}$, into harmonic oscillations to obtain
\begin{equation}
H_l(\xi) = \sqrt{2} \e^{\frac{l}{2}\left(\log (2l) -1\right)} \e^{\frac{\xi^2}{2}} \cos\left(\sqrt{2l} \xi - l\frac{\pi}{2}\right)
\end{equation}
which holds true with a relative error $\eta$ up to $\sqrt{4l\eta}$ (estimated from higher orders of the Taylor expansion). We fix $\eta=5\%$ such that the asymptotic form \eqref{eq:psi_asymptotic} is valid for $|\xi| < \sqrt{l/5}$.

\subsection{Derivation of the semiclassical vertex}
\label{app:B2}
We derive the semiclassical vertex, leading to \eqref{eq:H_HK_in_LL}, by assuming that the asymptotic form of the LLs $\psi_l \rightarrow \psi_l^\infty$ holds inside the entire integration region of the vertex. A basic calculation leads to 
\begin{widetext}
\begin{align}
\mac{V}_{ijkl}^{\nu}(0) =& \int_{-\nu}^{\nu} \d\xi_1 \d \xi_2 \d \xi_3 
 \psi_{i}^\infty\left(\xi_1\right)
 \psi_{j}^\infty\left(\xi_2\right)
 \psi_{k}^\infty\left(\xi_3\right)
 \psi_{l}^\infty\left(\xi_1-\xi_2+\xi_3\right)
\nonumber \\
=& \frac{\nu^3}{8\pi^2 (ijkl)^{1/4}} \sum_{\pm_{(i,j,k,l)}} \e^{-\ii \frac{\pi}{2}\left(\pm_i i\pm_j j\pm_k k\pm_l l\right)} \int_{-\nu}^{\nu} \d\xi_1 \d \xi_2 \d \xi_3 \e^{\ii \xi_1\left(\pm_i\sqrt{2i}\pm_l\sqrt{2l}\right)} \e^{\ii \xi_2\left(\pm_j\sqrt{2j}\mp_l\sqrt{2l}\right)}
\e^{\ii \xi_3\left(\pm_k\sqrt{2k}\pm_l\sqrt{2l}\right)}
\nonumber \\
=&\frac{\pi}{(ijkl)^{1/4}} \sum_{\pm_{(i,j,k,l)}} \e^{-\ii \frac{\pi}{2}\left(\pm_i i\pm_j j\pm_k k\pm_l l\right)}
\delta_{1/\nu}\left(\pm_i\sqrt{i/2}\pm_l\sqrt{l/2}\right)
\delta_{1/\nu}\left(\pm_j\sqrt{j/2}\mp_l\sqrt{l/2}\right)
\delta_{1/\nu}\left(\pm_k\sqrt{2k}\pm_l\sqrt{2l}\right)
\label{eq:app:vertex_semiclassical}
\\
\overset{\small i,j,k,l \gg 1}{\approx}& \frac{2\pi}{(ijkl)^{1/4}} \delta_{1/\nu}\left(\sqrt{2i}-\sqrt{2l}\right)
\delta_{1/\nu}\left(-\sqrt{2j}+\sqrt{2l}\right)
\delta_{1/\nu}\left(\sqrt{2k}-\sqrt{2l}\right) \cos\left(\frac{\pi}{2}\left(i-j+k-l\right)\right)
\end{align}
\end{widetext}
where
\begin{equation}
\delta_{1/\nu}(x) = \frac{\nu}{\pi} \frac{\sin(\nu x)}{\nu x} = \frac{1}{2\pi} \int_{-\nu}^\nu \d \xi \e^{\ii x \xi}
\end{equation}
and the sum extends over the 16 terms arising from different combinations of the signs. We refer to \eqref{eq:app:vertex_semiclassical} as the semiclassical vertex. 

In the limit where $i,j,k,l \gg 1$ only the sum combinations (upper,lower,upper,lower)-sign and (lower,upper,lower,upper)-sign remain relevant and $\delta_{1/\nu}$ become effectively $\delta$-functions. The vertex is hence diagonal in the LLs $\mac{V}_{ijkl}^{\nu} \propto \delta_{i,j,k,l}$. When normalizing the asymptotic wavefunction inside the integration interval $\mac{N}^2_l = \int_{-\nu}^\nu |\psi_l^\infty(\xi)|^2 \d \xi$ the entire prefactor for the interaction $\frac{\ell_B}{L} \mac{V}_{llll}^{L/(2\ell_B)}(k_x)/\mac{N}_l^4 = 1$ approaches 1. This leads to an effective HK-Hamiltonian in the LL basis, i.e. \eqref{eq:H_HK_in_LL} in the semiclassical limit.

\section{Calculation of the LL vertex \texorpdfstring{$V_{ijkl}$}{Vijkl}}
\label{app:vertex_exact_integrals}
\subsection{Introduction}
In the limit $L \gg \ell_B$ the integral $\mac{V}_{ijkl}^{L/(2\ell_B)}(k_x)$ can be solved exactly for all indices. The only important approximation for this limit is the extension of the integration boundary to infinity $L/(2\ell_B) \rightarrow \infty$. All dependencies on $k_x$ cancel out.

The vertex can be split up into two equivalent integrals
\begin{align}
V_{ijkl} &= \mac{V}_{ijkl}^\infty(k_x)
&= \int_{-\infty}^{\infty} \d z I_{ij}(z) I_{kl}(-z) 
\end{align}
where 
\begin{align}
I_{ij}(z) = \int_{-\infty}^{\infty} \d x \psi_i(x) \psi_j(x+z)
\end{align}
\subsection{Properties of \texorpdfstring{$I_{ij}$}{Iij}}
Here, we list some useful properties of $I_{ij}$
\begin{subequations}
\begin{align}
I_{ij}(0) &= \delta_{ij} \\
I_{ij}(z \rightarrow \infty) &= 0 \\
I_{ij}(-z) &= (-1)^{i+j} I_{ij}(z) \\
I_{ji}(z) &= (-1)^{i+j} I_{ij}(z)
\label{eq:app:I_prop4}
\end{align}
\end{subequations}
which can be easily shown by using the properties of $\psi_l$. Most importantly the integral $I_{ij}$ can be solved exactly, the solution is   
\begin{equation}
I_{ij}(z) = \e^{-z^2/4} z^{j-i} \sqrt{\frac{i!}{j!2^{j-i}}} \binom{j}{i} {}_1F_1(-i,1+j-i, z^2/2)
\label{eq:app_Iij_result}
\end{equation}
for $j\geq i$ where ${}_1F_1$ is Kummer's (confluent hypergeometric) function of the first kind \cite{Kummer1837}
\begin{equation}
_1F_1(\alpha,\beta, z) = \sum_{n=0}^\infty \frac{(\alpha+n-1)!}{(\alpha-1)!}\frac{(\beta-1)!}{(\beta+n-1)!} \frac{z^n}{n!}.
\label{eq:app_1F1}
\end{equation}
Whereas this is in general not a helpful representation, we emphasize that in the above equation ${}_1F_1$ is a sum over $i$ terms and hence a polynomial in $z$. 

\eqref{eq:app_Iij_result} is derived below w.l.o.g for $j\geq i$ (see \eqref{eq:app:I_prop4}):
\begin{widetext}
\begin{align}
I_{ij}(z) =& \int \d y \psi_{i}(y-z/2)\psi_j(y+z/2)
\\
=& \frac{\e^{-z^2/4}}{\sqrt{\pi 2^{i+j}i! j!}} \int \d y \e^{-y^2} H_i(y-z/2) \underbrace{H_j(y+z/2)}_{\sum_{k=0}^j \binom{j}{k} H_k(y) z^{j-k}}
\\
=& \frac{\e^{-z^2/4}}{\sqrt{\pi 2^{i+j}i! j!}}  \sum_{k,k'=0}^{i,j} \binom{i}{k} \binom{j}{k'} (-z)^{i-k} z^{j-k'} \underbrace{\int \d y \e^{-y^2} H_{k}(y)  H_{k'}(y)}_{2^k k! \sqrt{\pi}\delta_{k,k'}}
\\
=& \frac{\e^{-z^2/4}}{\sqrt{ 2^{i+j}i! j!}} 2^i z^{j-i} i!  \sum_{k=0}^{i} \frac{(-1)^k}{2^k k!} \binom{j}{i-k}  z^{2k}
\\
=& \e^{-z^2/4} z^{j-i} \sqrt{\frac{i!}{j!2^{j-i}}} \binom{j}{i} {}_1F_1(-i,1+j-i, z^2/2)
\end{align}
\end{widetext}

\subsection{Properties of \texorpdfstring{$V_{ijkl}$}{Vijkl}}
Here, we list some useful properties of $V_{ijkl}$:
First, half of the integrals evaluate to 0 due to an odd integrand
\begin{equation}
V_{ijkl} = 0 \quad \text{for } i+j+k+l \text{ odd}.
\end{equation}
The permutative relations 
\begin{subequations}
\begin{align}
V_{jikl} &= (-1)^{i+j} V_{ijkl}
\label{eq:app_V_prop1}\\
V_{kjil} &= V_{ijkl} \\
V_{ljki} &= (-1)^{i+l} V_{ijkl} \\
V_{ikjl} &= (-1)^{j+k} V_{ijkl} \\
V_{ilkj} &= V_{ijkl} \\
V_{ijlk} &= (-1)^{k+l} V_{ijkl}
\label{eq:app_V_prop6}
\end{align}
\end{subequations}
reduce the the number of independent tensor entries. 

\begin{widetext}
Most importantly, the integral can be evaluated exactly, the solution is for $j\geq i$  and  $l \geq k$
\begin{align}
    V_{ijkl} =& (-1)^{k+l} \sqrt{\frac{i!k!}{j!l!}} \begin{pmatrix} j\\i\end{pmatrix} \begin{pmatrix} l\\k\end{pmatrix}  {}_1F_1\left(-i,1+j-i;-\frac{\d}{\d c}\right)  {}_1F_1\left(-k,1+l-k;-\frac{\d}{\d c}\right) \left(-\frac{\d}{\d c}\right)^{(j-i+l-k)/2} \sqrt{\frac{2\pi}{c}} \Bigg |_{c=1} \nonumber\\
    =& \sqrt{2\pi} (-1)^{k+l} \sqrt{\frac{i! k!}{j! l!}} \sum_{n,n'=0}^{i,k} \frac{(-1)^{n+n'}}{n! n'!} \binom{j}{i-n} \binom{l}{k-n'} \frac{(2n+2n' + j-i+l-k-1)!!}{2^{n+n'+(j-i+l-k)/2}}.
\label{eq:app_Vijkl}
\end{align}
Note that $_1F_1$ are finite polynomials and that $j-i+l-k$ is even if and only if $i+j+k+l$ is even (if odd $V_{ijkl}=0$). By $\left(-\frac{\d}{\d c}\right)^n$ we mean $(-1)^n \frac{\d^n}{\d c^n}$ (the entire differential operator needs to be calculated first) and for calculation we may use that $(-1)^n \frac{\d^n}{\d c^n} \frac{1}{\sqrt{c}}= \frac{(2n-1)!!}{2^n}$ where the double factorial $!!$ denotes a factorial over all numbers with the same parity.

For some indices \eqref{eq:app_Vijkl} evaluates to simpler results. For equal indices $V_{iikk} = \sqrt{2\pi} L_i\left(-\frac{\d}{\d c}\right) L_k\left(-\frac{\d}{\d c}\right) \frac{1}{\sqrt{c}} \Big |_{c=1}$ where $L_k(x)$ are the Laguerre polynomials. This form simplifies to $V_{00ll} = \sqrt{2} \frac{\Gamma(l+1/2)}{\Gamma(l+1)}$ if one of the indices is 0. This result can be used to obtain an estimate of the scaling of the long range interaction between the LLs, since $V_{00ll} \rightarrow \sqrt{\frac{2}{l}}$ for $l\gg 1$.

From \eqref{eq:app_Vijkl} each matrix element of $V_{ijkl}$ can be calculated exactly be evaluating the finite sums. The numerical complexity increases for increasing indices. In practice one has to be careful when performing the sums. The summands have different signs and each of them is larger (in terms of its absolute value) then the total sum. This renders all summands relevant and requires arbitrary precision floating point operations from the numerical side. 

\eqref{eq:app_Vijkl} is derived below w.l.o.g for $j\geq i$  and  $l \geq k$ (see \eqref{eq:app_V_prop1}-(\ref{eq:app_V_prop6})):
\begin{align}
V_{ijkl} =& (-1)^{k+l} \int \d z I_{ij}(z) I_{kl}(z)
\\
=& (-1)^{k+l} \sqrt{\frac{i! k!}{j!l!}} \binom{j}{i} \binom{l}{k} \int_{-\infty}^\infty \d z \e^{-z^2/2} \left(\frac{z^2}{2}\right)^{\frac{j-i+l-k}{2}}  \underbrace{{}_1F_1(-i,1+j-i, z^2/2) {}_1F_1(-k,1+l-k, z^2/2)}_{\text{polynomials}}
\\
=& (-1)^{k+l} \sqrt{\frac{i! k!}{j!l!}} \binom{j}{i} \binom{l}{k} \left(-\frac{\d}{\d c}\right)^{\frac{j-i+l-k}{2}} {}_1F_1\left(-i,1+j-i, -\frac{\d}{\d c}\right) {}_1F_1\left(-k,1+l-k, -\frac{\d}{\d c}\right) \int_{-\infty}^\infty \d z \e^{-c z^2/2}\Bigg |_{c=1}   
\\
=& (-1)^{k+l} \sqrt{\frac{i! k!}{j!l!}} \binom{j}{i} \binom{l}{k} \left(-\frac{\d}{\d c}\right)^{\frac{j-i+l-k}{2}} {}_1F_1\left(-i,1+j-i, -\frac{\d}{\d c}\right) {}_1F_1\left(-k,1+l-k, -\frac{\d}{\d c}\right) \sqrt{\frac{2\pi}{c}}\Bigg |_{c=1}
\\
\overset{\text{\small (\ref{eq:app_1F1})}}{=}& (-1)^{k+l} \sqrt{\frac{i! k!}{j!l!}} \sum_{n,n'=0}^{i,k} \frac{(-1)^{n+n'}}{n! n'!} \binom{j}{i-n} \binom{l}{k-n'} \left(-\frac{\d}{\d c}\right)^{n+n'+\frac{j-i+l-k}{2}} \sqrt{\frac{2\pi}{c}}\Bigg |_{c=1}
\\
=& \sqrt{2\pi} (-1)^{k+l} \sqrt{\frac{i! k!}{j! l!}} \sum_{n,n'=0}^{i,k} \frac{(-1)^{n+n'}}{n! n'!} \binom{j}{i-n} \binom{l}{k-n'} \frac{(2n+2n' + j-i+l-k-1)!!}{2^{n+n'+(j-i+l-k)/2}}
\end{align}
\end{widetext}

\section{Short-time Fourier transformation}
\label{app:stft}
The STFT is a method from Fourier analysis to determine phase and frequency information for local sections of a signal changing over time. The basic idea is to perform several fast Fourier transformations of consecutive windows in the time domain to obtain the frequency for a segment in time. 

We will explain how typical STFT plots of oscillating functions with time dependant frequencies look like by considering a test function $g(t) = \exp \left(\ii f(t) t\right)$ where $f(t)$ is a slowly varying function with respect to $g$. We wish to evaluate the Fourier transform 
\begin{equation}
I(t_0,\omega) = \int_{-\infty}^{\infty} \e^{-\ii \omega t} g(t) w(t-t_0)
\end{equation}
as function of frequency $\omega$ and time $t_0$ and $w(t)$ is any windowing function which for proof of principle we choose to be a gaussian $w_\sigma(t) = \e^{-t^2/2/\sigma^2}$. The windowing function restricts the dominant part of integration region to $|t-t_0|/\sigma < 1$. The vague statement of $f$ being a slowly varying function can be formulated in more rigorous terms: First, for $|t-t_0|/\sigma < 1$ the Taylor expansion $f(t) = f(t_0) + f'(t_0) (t-t_0)$ holds. Secondly, the oscillations are fast with respect to the width of the window~$\nicefrac{\omega}{\sigma} \gg 1$. 

Under these assumptions, which are met in our MC data as well as in possible experimental data, the integral can be solved exactly by completing the square. The main result is that $I(t_0,\omega)$ is exponentially peaked at $\omega_{\max} = f(t_0) + t_0 f'(t_0)$. Therefore the STFT does not show $f(t)$ directly but only its linear approximation inside each segment. This can be used to efficiently reconstruct $f(t)$.

\section{LL interactions in the Hubbard model}
\label{app:Hubb}
Here, we provide details for the derivation of \eqref{eq:H_Hubb_LL} and derive the LL vertex for the Hubbard model $\tilde V_{ijkl}$.
\subsection{Obtaining the vertex}
We project the local Hubbard interaction in the LL basis ignoring its effects on the LL degeneracy. Hence, the calculation is similar to appendix~\ref{app:A} \eqref{eq:app:HK_derive_vertex}. We take $L/\ell_B \rightarrow \infty$ directly because the semiclassical limit is not of interest here. The interaction reads
\begin{align}
&\tilde{U} \sum_{y} c^\dag_{y,\ua} c_{y,\ua} c^\dag_{y,\da} c_{y,\da}
\nonumber\\
=&\frac{\tilde{U}}{\ell_B^2} \sum_{l_1,l_2,l_3,l_4} c^\dag_{l_1,\ua} c_{l_2,\ua} c^\dag_{l_3,\da} c_{l_4,\da} \int_{-\infty}^{\infty} \d y \nonumber\\
& \times  \psi_{l_1}\left(\frac{y}{\ell_B}+\ell_B k_x\right)
\psi_{l_2}\left(\frac{y}{\ell_B}+\ell_B k_x\right)  \nonumber\\
& \times \psi_{l_3}\left(\frac{y}{\ell_B}+\ell_B k_x\right) \psi_{l_4}\left(\frac{y}{\ell_B}+\ell_B k_x\right)
\nonumber\\
=& \frac{\tilde{U}}{\ell_B}\sum_{l_1,l_2,l_3,l_4} \tilde V_{l_1l_2l_3l_4} c^\dag_{l_1,\ua} c_{l_2,\ua} c^\dag_{l_3,\da} c_{l_4,\da} 
\label{eq:app:Hubb_derive_vertex}
\end{align}
where the vertex is
\begin{align}
\tilde V_{l_1l_2l_3l_4} =&  \int_{-\infty}^{\infty} \d\xi \psi_{l_1}\left(\xi\right)
\psi_{l_2}\left(\xi\right) \psi_{l_3}\left(\xi\right) \psi_{l_4}\left(\xi\right).
\label{eq:app:Vijkl_def_Hubb}
\end{align}

\subsection{Calculation of vertex \texorpdfstring{$\tilde V_{ijkl}$}{VHubbijkl}}
We evaluate the LL vertex $\tilde V_{ijkl}$ for the Hubbard model \eqref{eq:app:Vijkl_def_Hubb} exactly. From the properties of $\psi_l$ it is obvious that half of the entries are zero
\begin{align}
\tilde V_{ijkl} = 0 \quad \text{for } i+j+k+l \text{ odd}.
\end{align}
similar to the HK model. Furthermore, the vertex is symmetric in each index pair $\tilde V_{ijkl} = \tilde V_{jikl} = \tilde V_{kjil} = ...$

To solve the integral we use the series representation of the Hermite polynomials \cite{Abramowitz1972}
\begin{equation}
H_l(x) = l! \sum_{n=0}^{\lfloor{l/2}\rfloor} \frac{(-1)^n}{n!(l-2n)!} (2x)^{l-2n}
\end{equation}
where $\lfloor x\rfloor$ is the largest integer $\leq x$. For $i+j+k+l$ even the vertex is
\begin{widetext}
\begin{align}
\tilde V_{ijkl} =& \int_{-\infty}^\infty \d \xi \frac{1}{\pi \sqrt{2^{i+j+k+l}i!j!k!l!}} \e^{-2\xi^2}H_i(\xi)H_j(\xi)H_k(\xi)H_l(\xi) 
\\
=& \frac{1}{\pi} \sqrt{i!j!k!l!} \sum_{n_1,n_2,n_3,n_4=0}^{\lfloor{i/2}\rfloor,\lfloor{j/2}\rfloor,\lfloor{k/2}\rfloor,\lfloor{l/2}\rfloor}\frac{(-2)^{-n_1-n_2-n_3-n_4}}{n_1!n_2!n_3!n_4!(i-2n_1)!(j-2n_2)!(k-2n_3)!(l-2n_4)!}
\nonumber\\ &\times
\int_{-\infty}^\infty \d \xi (2 \xi^2)^{(i+j+k+l)/2-n_1-n_2-n_3-n_4} \e^{-2\xi^2}
\\
=& \frac{\sqrt{i!j!k!l!}}{\sqrt{2\pi}}  \sum_{n_1,n_2,n_3,n_4=0}^{\lfloor{i/2}\rfloor,\lfloor{j/2}\rfloor,\lfloor{k/2}\rfloor,\lfloor{l/2}\rfloor}\frac{(-2)^{-n_1-n_2-n_3-n_4}}{n_1!n_2!n_3!n_4!(i-2n_1)!(j-2n_2)!(k-2n_3)!(l-2n_4)!}
\nonumber\\ &\times
\left(-\frac{\d}{\d c}\right)^{(i+j+k+l)/2-n_1-n_2-n_3-n_4} \frac{1}{\sqrt{c}} \Bigg|_{c=1}
\\
=& \frac{1}{\sqrt{2\pi}} \sqrt{\frac{i!j!k!l!}{2^{i+j+k+l}}} \sum_{n_1,n_2,n_3,n_4=0}^{\lfloor{i/2}\rfloor,\lfloor{j/2}\rfloor,\lfloor{k/2}\rfloor,\lfloor{l/2}\rfloor}\frac{(-1)^{n_1+n_2+n_3+n_4} (i+j+k+l-2[n_1+n_2+n_3+n_4]-1)!!}{n_1!n_2!n_3!n_4!(i-2n_1)!(j-2n_2)!(k-2n_3)!(l-2n_4)!}
\end{align}
\end{widetext}
which is a series that can be computed exactly.

\subsection{QO in the HK model with fixed particle number}

\begin{figure}
    \centering
    \includegraphics[width=\columnwidth]{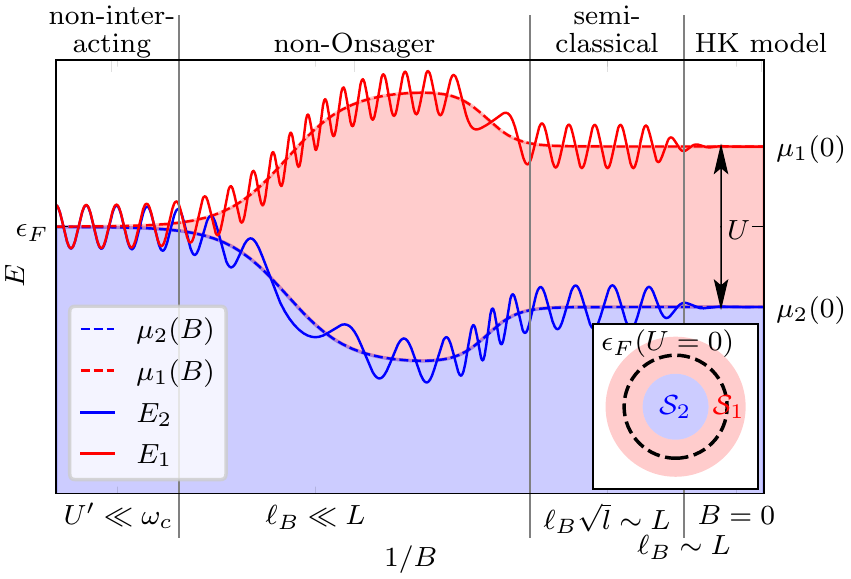}
    \caption{Schematic image of the DOS and the QO of e.g. the GS energy $E_1+E_2$ for fixed particle number in 2D, whereas Fig.~\ref{fig:fig1} is for fixed chemical potential. In the HK model at $B=0$ momentum states are double occupied up to $\mu_2(0)=\mu-U$ and single occupied from $\mu_2(0)$ to $\mu_1(0)=\epsilon_{F}$ where $\epsilon_{F}$ is the Fermi energy in the non-interacting limit. Due to the constant DOS in 2D the interaction leads to a symmetric single occupied region around the Fermi energy, see inset. At finite magnetic field the pseudo Fermi energies become magnetic field dependent, but such that the total particle number is conserved. Due to energetic constraints the drop of $\mu_2(0)$ in the non-Onsager regime will be less pronounced.}
    \label{fig:fig1N}
\end{figure}

In the main text, we have concentrated on results for a fixed chemical potential. In Fig.\ref{fig:fig1N} we show the schematic effect of keeping the particle number fixed which will introduce small quantiative changes. 

\clearpage
\bibliography{bib}

\end{document}